\documentclass[useAMS,usenatbib]{mn2e}

\usepackage{epsf}
\usepackage{url}

\title[Galaxy evolution during cluster merger]{The impact of a major cluster merger on galaxy evolution in MACS\,J0025.4-1225}
\author[C.-J. Ma, H. Ebeling, P. Marshall \& T. Schrabback]{C.-J. Ma$^{1}$, H. Ebeling$^{1}$, P. Marshall$^{2}$ \& T. Schrabback$^{3}$\\
$^{1}$Institute for Astronomy, University of Hawaii, 2680 Woodlawn Drive, Honolulu, HI 96822, USA \\
$^{2}$Kavli Institute for Particle Astrophysics and Cosmology, Stanford University, Stanford, CA 94305, USA \\
$^{3}$Leiden Observatory, Leiden University, Niels Bohrweg 2, NL-2333 CA Leiden, Netherlands \\
}
\begin{document}

\date{}

\pagerange{\pageref{firstpage}--\pageref{lastpage}} \pubyear{2010}

\maketitle

\label{firstpage}

\begin{abstract}
We present results of an extensive morphological, spectroscopic, and photometric study of the galaxy population of MACS\,J0025.4$-$1225 ($z=0.586$), a major cluster merger with clear segregation of dark and luminous matter, to examine the impact of mergers on galaxy evolution. Based on 436 galaxy spectra obtained with Keck DEIMOS, we identified 212 cluster members within 4 Mpc of the cluster centre, and classified them using three spectroscopic types; we find 111 absorption-line, 90 emission-line (including 23 e(a) and 11 e(b)), and 6 E+A galaxies. The fraction of absorption(emission)-line galaxies is a monotonically increasing(decreasing) function of both projected galaxy density and radial distance to the cluster center. More importantly, the 6 observed E+A cluster members are all located between the dark-matter peaks of the cluster and within $\sim 0.3$Mpc radius of the X-ray flux peak, unlike the E+A galaxies in other intermediate-redshift clusters which are usually found to avoid the core region. In addition, we use Hubble Space Telescope imaging to classify cluster members according to morphological type. We find the global fraction of spiral and lenticular galaxies in MACS\,J0025 to be among the highest observed to date in clusters at z$>0.5$. The observed E+A galaxies are found to be of lenticular type with Sersic indices of $\sim2$, boosting the local fraction of S0 to 70 per cent between the dark-matter peaks. Combing the results of our analysis of the spatial distribution, morphology, and spectroscopic features of the galaxy population, we propose that the starburst phase of these E+A galaxies was both initiated and terminated during the first core-passage about 0.5--1Gyr ago, and that their morphology has already been transformed into S0 due to ram pressure and/or tidal forces near the cluster core. By contrast, ongoing starbursts are observed  predominantly in infalling galaxies, and thus appears to be unrelated to the cluster merger. 

\end{abstract}


\section{Introduction}

It is well known that the galaxy population of clusters is dominated by passively evolving, red, early-type galaxies \citep[e.g.][]{dressler80, bower92, vandokkum98}, unlike the population of galaxies in less dense environments \citep[e.g.][]{gomez03, lewis02, cooper07}. Many cluster-specific environmental processes, such as ram-pressure stripping \citep{gunn72,bekki02}, strangulation\citep{larson80}, galaxy-galaxy harassment \citep{moore96}, or tidal disruption \citep{merritt83,byrd90} have been proposed to explain the effects of cluster assembly on the galaxy population. However, the picture of galaxies being accreted in a static cluster environment is overly simplistic.  Many optical and X-ray studies  \citep[e.g.][]{jones99, depropris04} reveal that a significant fraction, if not all, of the clusters even at recent epochs are still growing through mergers of galaxy groups and sub-clusters. To get a correct picture of the effects of environment on galaxy evolution, the dynamic processes governing the formation of these large-scale structures must be taken into account. For instance, the predominance of passively evolving galaxies in clusters was suggested to have its roots in the preprocessing of galaxies in the group environment \citep[e.g.][]{zabludoff98,kodama01, tran09,wilman09}, thus not being directly related to the much denser environment in clusters. 

Although environment has thus been identified to play a critical role in the evolution of galaxies, the effect of events that cause the most dramatic change in environment, cluster mergers, is still poorly understood \citep{girardi02}. For instance, it is still controversial whether star-formation activity would be enhanced during cluster mergers, as favored by numerical simulations \citep[e.g.][]{bekki02,kronberger08}. Observationally, this picture is supported by the classical studies of \citet{caldwell93} and \citet{caldwell97}, which find evidence of recent star bursts in merging clusters. More recently, \citet{owen05} suggested that the excess of radio-loud galaxies around the cluster Abell~2125 may be related to the ongoing cluster merger, and \citet{ferrari05} report on actively star-forming galaxies in the region between the two sub-clusters of the merging cluster Abell~3921 ($z=0.094$). In addition, \citet{hwang09} compare the two merging clusters Abell~168 and Abell~1750, and attribute the different amount of galaxy activity, including star formation and nuclear activity, in the region between the subclusters to the differences in the merger history of the two clusters. On the other hand, the ages of numerous post-starburst galaxies found in Abell~3921 are too old for the starbursts to have been triggered by the ongoing merger \citep{ferrari05}. Also, a 24$\mu$m observation of ClJ0152.7$-$1357 \citep[$z\sim0.83$, ][]{ebeling00} by \citet{marcillac07} suggests that most of the mid-infrared luminous galaxies are infalling and thus not related to the merging process. 

The cause of the conflicting observational evidence may lie in the complexity of the dynamics of cluster mergers. For instance, the specific phase in which a merger is observed, the viewing angle, and the relative masses of the subclusters can all be expected to have a direct effect on the triggering of star formation and the resulting spatial and spectral observational data \citep[see e.g.][]{hwang09}. Finding merging systems whose geometry and merger history are immediately evident is thus of paramount importance. To date, only three such systems have been identified through a clear segregation between dark matter and intra-cluster gas: the Bullet Cluster  \citep[1E0657-56, $z=0.296$;][]{clowe04,clowe06}, Abell~520 \citep[$z=0.201$;][]{mahdavi07}, and MACS\,J0025.4$-$1225 \citep[$z=0.586$, hereafter MACSJ0025;][]{bradac08}. Among these three systems, MACSJ0025 stands out as the best target for a study of the effects of a major cluster merger because the masses of the two sub-clusters are similar\footnote{Note, however, that the supersonic head-on collision of two systems of very different mass in the Bullet Cluster enabled a detailed study of the effects of ram pressure from supersonic gas during a cluster merger \citep[see][]{chung09}.}. 

In this paper, we will report spectroscopic and morphological results for galaxies around MACSJ0025, a cluster discovered in the MAssive Cluster Survey \citep{macs,ebeling07}, and, at L$_{X,Chandra} = 8.8\pm 0.2\times~10^{44}$~ergs~s$^{-1}$, one of the most X-ray luminous clusters at z$>0.5$ \citep{ebeling07}. Our paper is structured as follows. \S\ref{sec:dynamics} summarizes the dynamical structure of MACSJ0025 as studied by \citet{bradac08}. \S\ref{sec:data} gives an overview of the observation and data reduction procedures. \S\ref{sec:analysis_technique} describes the analysis of photometric, spectroscopic and morphological data. In \S\ref{sec:result}, we present our results. Finally, we discuss a simple merger model in \S\ref{sec:discussion} and provide a brief summary in \S\ref{sec:summary}.  Throughout this paper, we adopt the concordance $\Lambda$CDM cosmology with $h_0=0.7$, $\Omega_{\Lambda}=0.7$, $\Omega_m = 0.3$. Magnitudes are quoted in the AB system.

%\onecolumn
\begin{figure*}
\epsfxsize=0.8\textwidth
\epsffile{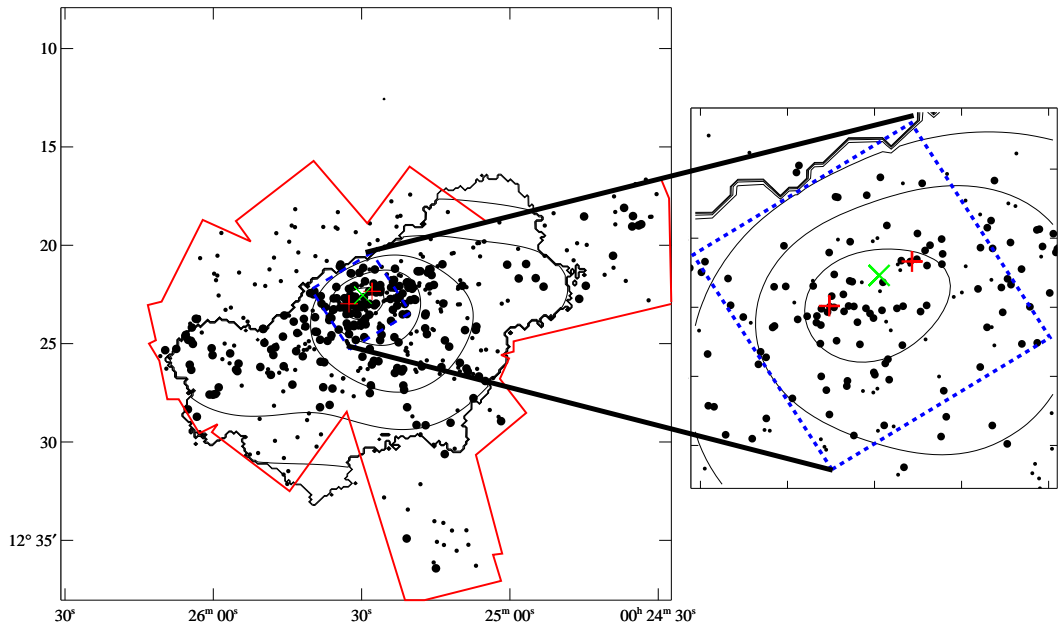}
 \caption[f1.eps]{Spatial distribution of galaxies
  observed spectroscopically: small dots mark the location of all sources with
  measured spectrum, large dots mark confirmed cluster members. The overlaid
  contours show the projected galaxy density as computed by
  \textit{asmooth} \citep[ ][ see \S\ref{sec:phot_galdens} for details]{asmooth}.  The red boundary
  indicates the outline of all DEIMOS masks and defines the study region for
  this paper. The blue dash square indicate the HST ACS image mentioned in \S\ref{sec:data_image_hst}, and defines the area for the morphology analysis in \S\ref{sec:morph}. The two plus signs show the location of gravitational lensing peaks \citep{bradac08}, and the cross sign shows the location of the X-ray flux peak.\label{plane}}
\end{figure*}
%\twocolumn

\section{Summary of dynamics of MACSJ0025}\label{sec:dynamics}

MACSJ0025 is the second cluster that shows a significant segregation between the hot gas traced by the X-ray surface brightness, and the mass traced by the gravitational shear \citep{jeyhan08,bradac08} in a simple two-body merger configuration.  We here briefly summarize the dynamics of the system as presented by \citet{bradac08}.

Like many other merging clusters, MACSJ0025 features a galaxy distribution in redshift space as well as in projection on the sky which, taken alone, does not show obvious evidence of substructure. Its galaxy redshift distribution is consistent with a Gaussian centered at z$_{cl}=0.5857$ with $\sigma_{cl} = 835^{+58}_{-59}$ km~s$^{-1}$ with only mild evidence of a bimodal distribution\footnote{In \S\ref{sec:data_spect_cat}, we discuss updates to the spectroscopic sample; the results are similar. } . In addition, the Dressler-Shectman test \citep{dstest} detects only marginal cluster substructure. The redshift difference between the two sub-clusters is small ($\Delta z=0.0005\pm0.0004$), which implies that the two subclusters are colliding almost in the plane of the sky (within $5\degr$). 

The X-ray temperature of the intracluster medium (ICM) within a circular region of radius $660$~kpc centered on the X-ray peak is $6.26^{+0.50}_{-0.41}$~keV; the metallicity of the ICM is $Z=0.37\pm0.10$ solar. The corresponding X-ray mass of the gas is $0.55\pm0.06 \times 10^{14}$~M$_{\sun}$. The virial radius for a cluster with the same gas temperature is then $\sim1.3$~Mpc, calculated following the prescription provided by \citet{arnaud02}. Finally, the radius at which a Milky-way-like galaxy would lose all its gas due to ram-pressure stripping is $\sim0.7$~Mpc (see \cite{ma08} for details). 

The gravitational lensing analysis of \citet{bradac08} shows the two peaks of the mass distribution being offset from the X-ray peak by $0.82\arcmin$ for the southeastern peak and $0.50\arcmin$ for the northwestern peak. Since the X-ray centre lies close to the projected line connecting the gravitational lensing peaks, the separation of the two lensing peaks is thus $1.32\arcmin$ ($0.52$~Mpc). The masses estimated from the lensing model are $2.5^{+1.0}_{-1.7}\times10^{14}$~M$_{\sun}$ and $2.6^{+0.5}_{-1.4}\times10^{14}$~M$_{\sun}$ for the southeastern and northwestern peak, respectively. 

\section{Observation and Data Reduction}\label{sec:data}

\subsection{Imaging data}\label{sec:data_image}
\subsubsection{Groundbased observations}\label{sec:data_image_phot} 

Our object catalogue is based on images in five optical bands (B, V, R$_c$, I$_c$, and z$^\prime$) obtained with the Suprime-Cam \citep{suprimecam} wide-field imager on the Subaru 8m telescope on Mauna Kea. We also include imaging data for a near-ultraviolet band (u$^*$) obtained with the MegaPrime camera, and two near-Infrared bands (J and K$_s$) obtained with the WIRCAM\citep{wircam} instrument, both on the CFHT 3.6m telescope on Mauna Kea. We use the same pipeline described in \citet{ma08, jeyhan08} to reduce the optical and u$^*$-band data. Near-infrared images are obtained by stacking the preprocessed data provided by CFHT using software\citep{scamp,swarp} developed by E. Bertin\footnote{\url{http://astromatic.iap.fr/software}}. The construction of a photometric object catalogue then follows the same procedure as \citet{ma08} and \citet{jeyhan08}. The photometry of the Suprime-Cam imaging data is calibrated using hundreds of stars with magnitudes ranging from 16 mag to 19 mag in a nearby SDSS field\footnote{The SDSS photometry is transformed into the Johnson-Cousin system using the equation of Lupton (2005) provided on \url{http://www.sdss.org/dr5/algorithms/sdssUBVRITransform.html.}}.  The calibration of our MegaCam data is supplied by CFHT's Elixir project\footnote{http://www.cfht.hawaii.edu/Instruments/Elixir/}. Finally, the 2MASS catalogue \citep{2mass}\footnote{This publication makes use of data products from the Two Micron All Sky Survey, which is a joint project of the University of Massachusetts and the Infrared Processing and Analysis Center/California Institute of Technology, funded by the National Aeronautics and Space Administration and the National Science Foundation.}
 is used to calibrate the photometry of the stacked WIRCAM images.

\subsubsection{HST imaging}\label{sec:data_image_hst}

The central $\sim1.3\times 1.3$Mpc$^2$ region of the cluster
      (Fig.~\ref{plane}) is covered by HST ACS images in the
      F555W filter (4140s) and in the F814W filter (4200s),
      obtained as part of the Cycle-13 proposal GO-10703 (PI
      Ebeling). The data were reduced using the HAGGLeS pipeline
      (Marshall, P. et al. 2009 in preparation), which employs
      MultiDrizzle\footnote{MultiDrizzle version 2.7.0} (Koekemoer et al. 2002)
      for the distortion correction, cosmic ray rejection, and stacking.
      Following manual masking of satellite trails, cosmic ray
      clusters, and scattered light in the flat-fielded frames,      
      we iteratively refine relative shifts and rotations
      by cross-correlating the positions of compact sources. 
      The individual
      frames in each filter were then drizzled together onto a common 0\farcs03 pixel grid
      using a square kernel with pixfrac 0.8, applying updated bad pixel masks and optimal weights
      (Schrabback et al. 2009).

\subsection{Spectroscopy}\label{sec:data_spect}
The spectroscopic data set used here is compiled primarily from observations of nine multi-object spectroscopy (MOS) masks with the DEIMOS spectrograph on the Keck-II telescope. The area covered by these masks is marked in Fig.~\ref{plane}. We also use 16 additional spectra obtained with the Low Resolution Image Spectrometer (LRIS) on the Keck-I telescope, and the Gemini Multi-Object Spectrograph (GMOS) on the Gemini telescope, of objects which meet our selection criteria (\S\ref{sec:data_spect_objsel}) and have not been observed with DEIMOS. 

Following the same strategy as \citet{ma08}, DEIMOS was used with the 600ZD grating, GC455 filter, and centre wavelength at 6700$\AA$ as a compromise between the requirements of wavelength coverage from the [OII] $\lambda\lambda 3727$ line to the H$_{\beta}$ line at a redshift of about 0.59, and our desire for at least moderately high spectral resolution. We integrated for $3\times1800$s for each mask; no flux calibration was performed. The data were reduced using the standard DEIMOS pipeline\footnote{The analysis pipeline used to reduce the DEIMOS data was developed at UC Berkeley with support from NSF grant AST-0071048.} developed by the DEEP2 team \citep{cooper07}. Redshifts were determined and verified manually using at least two prominent spectral features, such as (in absorption) Calcium H and K, H$_{\delta}$, or the G band, and (in emission) [OII] $\lambda\lambda 3727$, H$_{\beta}$, or [OIII] $\lambda\lambda 4959,5007$. 

%%%
\begin{figure*}
    \epsfxsize=0.8\textwidth
    \epsffile{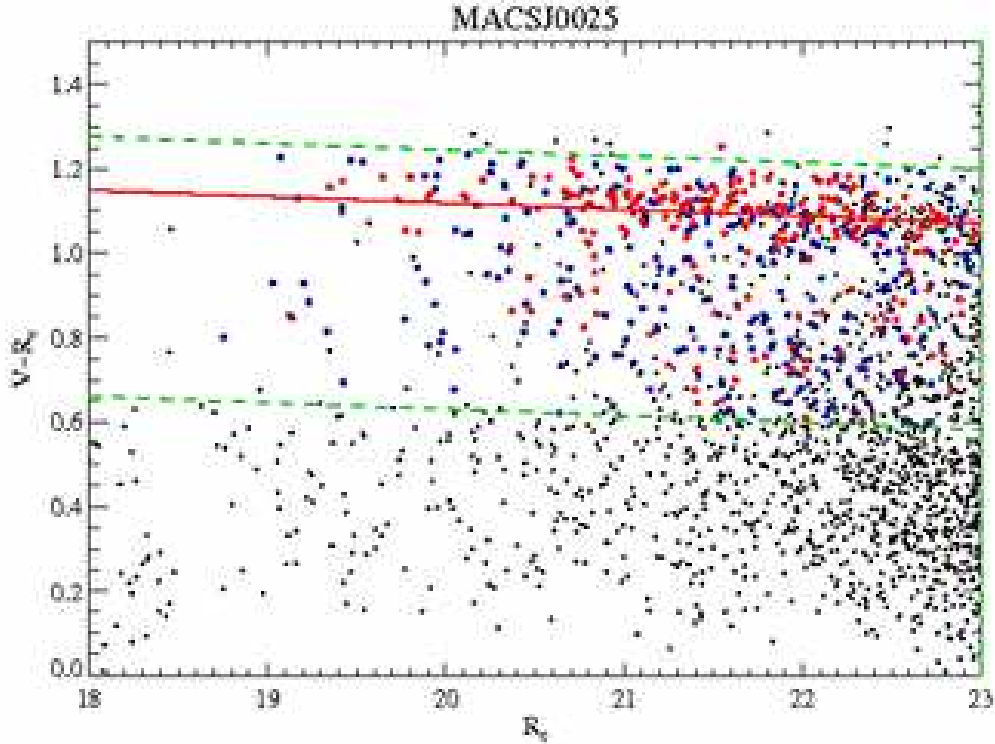} 
    \caption[f2.eps]{Color magnitude diagram
    for galaxies inside our study region as defined in Fig~\ref{plane}. Blue
    filled circles mark objects with observed spectra; red filled circles denote
    spectroscopically confirmed cluster members (cf.\
    Fig.~\ref{redshiftplt}). Black dots are all remaining objects in our
    SExtractor galaxy catalogue. The dashed lines indicate the selection
    criteria used for our spectroscopic survey. The red sequence marked by the
    solid line is obtained from a linear fit to the data points redder than
    V-R$_{c}\sim 0.9$ and brighter than R$_{c}\sim 23.0$. \label{cmd}}
\end{figure*}

\subsubsection{Object Selection and Completeness}\label{sec:data_spect_objsel}

In order to maximize the efficiency of cluster member detections while minimizing potential selection bias, we chose our spectroscopic targets following the approach taken by \citet{ma08}. Since, at the beginning of this project, imaging data in too few passbands were available to determine credible photometric redshifts, targets for spectroscopic follow-up observations were selected using the color magnitude diagram for the V and R$_c$ filters (Fig. ~\ref{cmd}). 
To maximize completeness, we limited our survey to relatively bright objects ($m_{\rm R_c}\leq 23.0$, which is, equivalently, M$_{\rm Rc} \leq $M$^*$+2.7 at z=$0.59$) while adopting a generous color cut around the cluster red sequence, extended toward the blue end of the distribution. We used our catalogue of photometric redshifts (\S\ref{sec:phot_phz}) to compute the fraction of missed blue cluster members($2.3\%$). We note that incompleteness of our spectroscopic survey is severely color-dependent which, as we will show in \S\ref{sec:spect_cmd}, could potentially cause the fraction of starburst galaxies, but not the fraction of E+A galaxies, to be underestimated. 

\begin{figure}
\epsfxsize=0.47\textwidth 
\epsffile{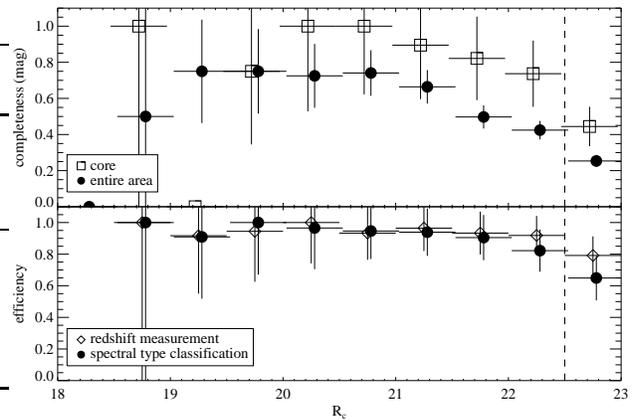}
\caption[f3.eps]{Completeness (top) and efficiency (bottom)
  of our spectroscopic survey as a function of galaxy magnitude. The color
  criterion of Fig.~\ref{cmd} has been applied. Top: survey completeness for the
  entire study region (filled symbols) and for the cluster centre region with radius $1~Mpc$ (open symbols). Bottom:
  survey efficiency, i.e. success rate for redshift determination (open symbols)
  and spectral classification (filled symbols). In both panels, the dashed line
  indicates the magnitude limit of the spectroscopic
  sample.  \label{completeness}}
\end{figure}

We define the completeness of our spectroscopic survey as the ratio of the
number of sources observed to the number of sources in the photometric catalogue that meet the selection criteria (i.e. that fall within the color band marked in Fig.~\ref{cmd} and exceed the magnitude limit), including spectra from which we failed to measure a credible redshift. In addition, we define the efficiency of our survey as the fraction of the number of high-quality spectra from which both the redshift and the equivalent width of spectral features can be measured accurately (again
within the color limits marked in Fig.~\ref{cmd}). The completeness and
efficiency as a function of galaxy magnitude are shown in the top panel of
Fig.~\ref{completeness}.  Inside the entire study region, the completeness is roughly constant at about $75\%$ at R$_c< 21.5$, with the curve only dropping below $40\%$ at the very last bin before the magnitude limit at R$_c=23.0$. Also shown in the top panel of Fig.~\ref{completeness} is the completeness within a radius of 1 ~Mpc from the centre of the cluster: the completeness is above $80\%$ at R$_c<22.5$\footnote{The completeness may drop because of small number statistics at the brightest magnitudes.}, and drops to $40\%$ at the last bin. Because of  the significantly reduced completeness at the faint end of our survey, we adopted a more restrictive magnitude limit (R$_c<22.5$, shown as the dashed lines in Fig.~\ref{completeness}.) for the statistical analysis of the properties of different spectral types of galaxies (\S\ref{sec:spect_spatdistr}) to prevent a potential bias caused by the luminosity dependence of the spectral type distribution. For all other analyses, however, we used the original magnitude limit, since the statistics in the faintest magnitude bin (28 cluster members out of 60 galaxies with spectroscopic data) are sufficient for our purposes. The differences between the completeness curves shown in Fig.~\ref{completeness} (top) demonstrate that the completeness of our survey is not spatially uniform; the {\em relative} spatial distribution of galaxies of different types should, however, be unaffected.

The bottom panel in Fig.~\ref{completeness} shows our efficiency in determining the galaxy redshift as well as a secure spectroscopic classification. The criteria used to classify galaxies according to spectral type (defined in the following two sections) are much more demanding than those required for a simple redshift measurement; still, the efficiency for spectroscopic classification remains about $0.9$ at all magnitudes except for the final bin, reflecting the good signal-to-noise ratio of most of the spectra. About $10\%$ of our high-quality spectra yield spectroscopic redshifts but do not allow a spectral type to be assigned; this is usually caused by the key spectral features falling on the chip gap or coinciding with telluric lines. At the faintest magnitudes probed by this work, the spectral classification begins to be limited by the uncertainty of the equivalent-width measurements, although we are still  able to classify about $70\%$ of the respective subsample.

\subsubsection{Spectroscopic catalogue and galaxy redshift distribution}\label{sec:data_spect_cat}

\begin{figure}
\epsfxsize=0.47\textwidth
\epsffile{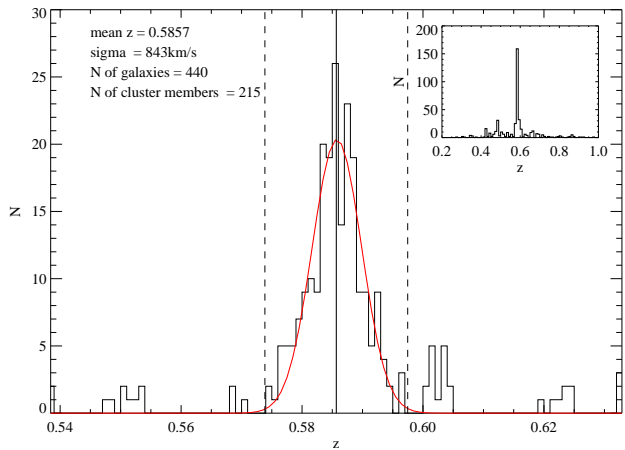} \caption[f4.eps]{Redshift
  distribution for all spectroscopically observed galaxies. The systemic
  redshift of MACSJ0025 is found to be $z_{\rm cl}=0.5857$. The dashed
  lines at $z=0.574$ and $z=0.597$ ($z=z_{\rm cl}\pm3\sigma$) mark the
  spectroscopic extent of the cluster.  A zoomed-out view of the full redshift range of the galaxies observed by us is
  shown in the sub-panel.\label{redshiftplt}}
\end{figure}

Much of the spectroscopic galaxy sample compiled for this work was used before in the study of \citet{bradac08} to perform a basic dynamical analysis. Since then, we have doubled the number of galaxies with spectroscopic data; the resulting final spectroscopic catalogue comprises 436 galaxies, of which 208 are cluster members (Table~\ref{spectsum}). Fig.~\ref{plane} shows the distribution of the spectroscopically observed galaxies within our study region which is highlighted by the outline of the DEIMOS masks. The full redshift distribution, shown in Fig.~\ref{redshiftplt}, is consistent with the one presented in Fig.~1 of \citet{bradac08} and is well described a single Gaussian. In addition, the systemic cluster redshift (z$_{cl}=0.5857$) and the velocity dispersion ($\sigma_{cl}=843$ km s$^{-1}$) of the 116 cluster members within the virial radius are also consistent with the values reported in \citet{bradac08} and \citet{ebeling07}. We therefore assign cluster membership to all galaxies with \mbox{$0.574<z<0.597$}, the redshift range given by $z=z_{cl}\pm 3\sigma$.

%Table 1 : spectral type summary
\begin{table}
\centering
 \caption{Number statistics of spectral types\label{spectsum}}
 \begin{tabular}{p{2cm}cccc}

% \tabletypesize{\scriptsize}
% \tablewidth{0pc}
% \tablecolumns{5} 

\hline 
Type & \multicolumn{2}{c}{All}                             & \multicolumn{2}{c}{cluster members}   \\
         &   R$_c<23.0$  & R$_c<22.5$                 &  R$_c<23.0$  & R$_c<22.5$  \\
\hline

redshift measurements                  &  436   & 383  &  212  &  182  \\
spectral type classification            &  421   & 375  &  207  &  180  \\
\cline{1-5}  \vspace*{-2mm} \\
Emission-line galaxies                  &  236   &  204 &    90   &  75   \\
Absorption-line galaxies               &  171   &  160  & 111   & 102   \\
E+A galaxies                                   &  14     &    11  &     6   &  3       \\  
\cline{1-5} \vspace*{-2mm} \\
e(a)                                                    & 57      & 53    &   23    &  22  \\
e(b)                                                    & 38      & 26    &   11    &   6   \\
e(c)                                                    &  141   &  125 &   56    &   47 \\
\hline
\end{tabular}
\end{table}

\begin{figure}
\epsfxsize=0.47\textwidth 
\epsffile{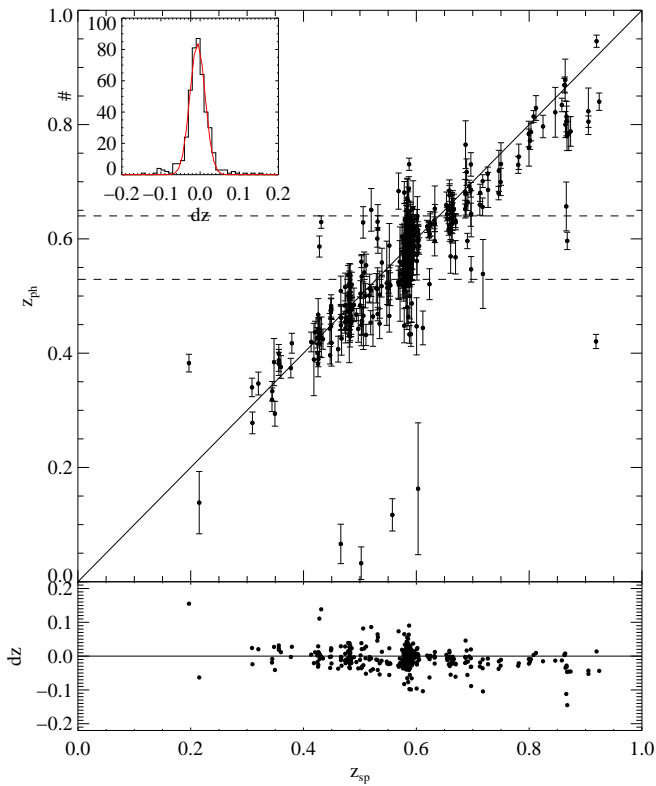}
\caption[f5.eps]{Photometric versus spectroscopic redshifts for
  galaxies in the field of MACSJ0025.  The lower panel shows
  the residuals d$z{=}z_{\rm phot}-z_{\rm spec}$ as a function of the spectroscopic redshift, while the
  insert in the top left corner shows the histogram of the residuals as well as
  a Gaussian model from which we derive $\Delta z{=}0.019$ as an estimate of the uncertainty of the photometric redshifts. \label{pzVSspz}}
\end{figure}

%Table 2: summary of definition
\begin{table}
  \centering
  \caption{Definition of spectral types\label{spectdef}}
  \begin{tabular}{cp{5.5cm}}
% \tabletypesize{\scriptsize}
% \tablewidth{0pc}
% \tablecolumns{2} 
  \hline 
  Type & Criteria \\
  \hline  

Emission-line     & detection of [OII] or $H_{\beta,em}$   \\
Absorption-line  & no detection of [OII] and $H_{\beta,em}$, and EW($H_{\delta}$)${<}4$\AA \\
E+A                    & no detection of [OII] and $H_{\beta,em}$, and EW($H_{\delta}$)${>}4$\AA \\
\cline{1-2}\vspace*{-2mm} \\
e(a)             & Emission-line galaxies with EW($H_{\delta}$)$>4$\AA \\
e(b)             & Emission-line galaxies with EW($H_{\delta}$)$<4$\AA\, and EW([OII])$<-25$\AA \\
e(c)             & Emission-line galaxies with EW($H_{\delta}$)$<4$\AA\, and EW([OII])$>-25$\AA \\
\hline
\end{tabular}
\end{table}

\section{Analysis technique}\label{sec:analysis_technique}

\subsection{Photometric redshift}\label{sec:phot_phz}

To derive photometric redshifts for the galaxies with \mbox{m$_{\rm R_c}<$24.2} (\mbox{$\sim$~M$^*$+4.0 at z$=$0.59}), we use the adaptive SED-fitting code \textit{Le Phare}\footnote{\url{http://www.cfht.hawaii.edu/~arnouts/LEPHARE/cfht\_lephare/lephare.html}} \citep{arnouts99,ilbert06,ilbert09}. In addition to its ${\chi}^2$ optimization during SED fitting, \textit{Le Phare} adaptively adjusts the photometric zero points by using the sample with spectroscopic redshifts as a training set\footnote{The resulting adjustments, which reflect the systematic uncertainty of the absolute photometric calibration as well as the complexity of matching SED templates to the observed spectra, for all bands are u$^*$:$0.081$, B:$0.066$, V:$0.090$, R$_c$:$-0.016$, I$_c$:$0.025$, z$^{\prime}$:$-0.091$, J:$-0.181$ and K$_s$:$0.102$, comparable to the values quoted by \citet{ilbert09}.}, which effectively reduces the fraction of catastrophic errors and also removed some systematic trends in the differences between the spectroscopic and photometric redshifts \citep{ilbert06}. We use the library of empirical templates from \citet{ilbert09}, which is included in the \textit{Le Phare} package. 
The training set is obtained from our database of spectroscopic redshifts by selecting all galaxies with an R$_c$-band magnitude between 20 and 22, and a minimal separation from the closest SExtractor-detected neighbour of $2\arcsec$ to avoid photometric errors from close pairs or blended sources. The resulting training set comprises 290 galaxies at redshifts ranging from 0.2 to 0.9. The comparison of spectroscopic and photometric redshifts is shown in Fig.~\ref{pzVSspz}. The statistical error of the photometric redshifts, as derived from the Gaussian distribution of the residuals shown in the insert of Fig.~\ref{pzVSspz}, is found to be $\Delta z = 0.019$.

Based on the photometric redshifts thus computed, we define cluster members to be all galaxies with \mbox{$| z_{ph} - z_{cl}| < 2 \sigma_{ph-z}$}, where \mbox{$\sigma_{ph-z} = (1+z_{cl}) \Delta z$}. Note that this redshift cut (\mbox{$\pm 2\sigma_{ph-z} = \pm0.061$}) is much more generous than the one used before for the spectroscopic redshifts ($\pm3\sigma = \pm 0.012$); the chosen cutoff value represents a good compromise between completeness and contamination. The resulting catalogue of ``photometric'' cluster members includes 1602 galaxies within the entire region imaged with SuprimeCam and allows us to estimate the local galaxy density within and outside our spectroscopic study area.

\begin{figure*}
  \epsfxsize=0.9\textwidth
  \epsffile{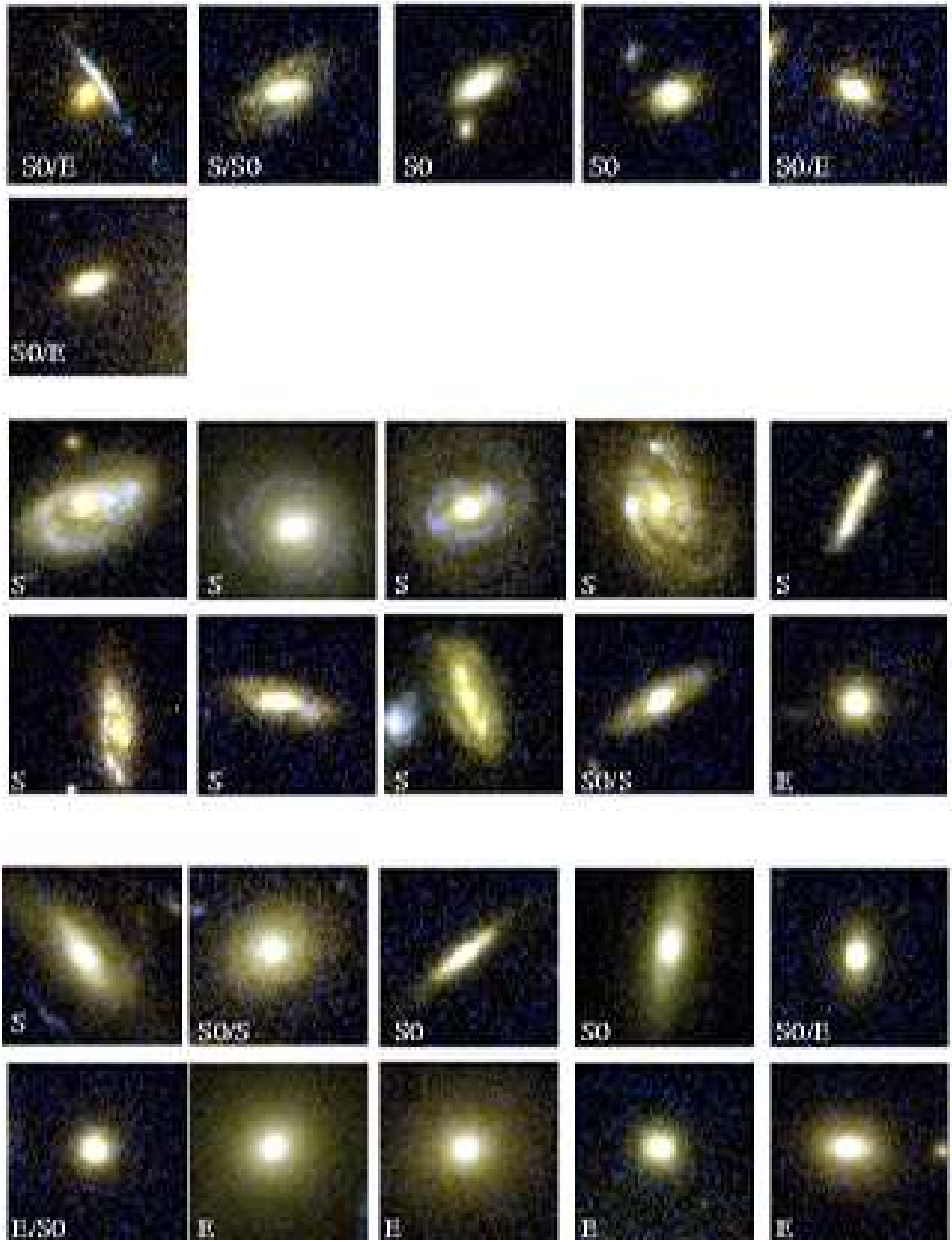}    
  \caption[f6]{ACS images (F505W and F814W) of selected cluster galaxies with stamp size $4.0\arcsec$. The six E+A cluster members are shown in the top two rows, and randomly selected emission-line and absorption-line galaxies in the cluster are shown in the middle and bottom two rows respectively. Note that the first E+A galaxies on the top row may be contaminated by a blue lensing arc. \label{example_stamps}}
\end{figure*}

\subsection{Projected galaxy density}\label{sec:phot_galdens}

Using the photometrically selected set of cluster members, we calculate the projected galaxy density in two ways. For the convenience of visual demonstration, a smoothed version of the spatial distribution of cluster members (shown in Fig.~\ref{plane}) is obtained with the adaptive-smoothing algorithm \textit{asmooth} \citep{asmooth} which preserves the signal-to-noise ratio of all significant features on all scales. In addition, in order to allow a direct comparison with the results from earlier work on the impact of cluster environment on galaxy activity \citep{kodama01, balogh04, sato06, ma08}, we also calculate, for all``photometric'' cluster members, the projected local density $\Sigma_{10}$, defined as the galaxy number density within a circle whose radius is given by the separation from the $10^{th}$ closest neighbour. The densities obtained from the two approaches are consistent with each other.

\subsection{Spectroscopic analysis}\label{sec:spect}

We perform equivalent-width (EW) measurements and spectral type classification following the prescription given in \citet{ma08}\footnote{The only difference is that we use the H$_{\delta}$ absorption line, rather than both H$_{\delta}$ and H$_{\gamma}$ as in \citet{ma08}, to classify E+A galaxies, since, for galaxies at the redshift of MACSJ0025, H$_{\gamma}$ falls onto the strong O$_2$ absorption feature of the atmosphere.}. In short, we classify galaxies into three major spectral types (absorption-line, emission-line, and E+A galaxies), and three sub-types (e(a), e(b), and e(c)) \citep[see also ][ for the properties of different spectral types]{dressler99,poggianti99,poggianti00,tran03,goto04}.  Specifically, E+A, also called ``post-starburst", galaxies are identified by the strong Balmer absorption lines indicating significant amount of young stars, and the absence of emission-lines indicating current star-formation activity has been terminated. On the other hand, e(a), also called ``dusty starburst", galaxies are mostly useful to trace the starburst galaxies. Their strong $H_{\delta}$ absorption lines indicate the recent burst of many young stellar population, and, in contrast to the E+A galaxies, the moderate [OII] emission exhibits the starburst has not been terminated. Furthermore, the infrared study of \citet{poggianti00} finds about half of the very luminous infrared galaxies [log(L$_{IR}$/L$_{\sun}$)$>$11.5] in their sample to exhibit e(a) spectra, which suggests that their moderate [OII] emission is heavily extinguished by dust. Also, the lack of e(b) spectra in the same sample indicates that e(b) characteristics are less well suited to identify starburst galaxies, although they exhibit stronger [OII] emission. For clarity, the definition of the spectral types is summarized in Table~\ref{spectdef}. 

The number of galaxies of each type is given in Table~\ref{spectsum}. Note that the fraction of E+A galaxies in this cluster is $\sim2.9\pm1.2\%$ for $m_{\rm R_c}{<}23.0$, and $\sim2.2\pm1.1\%$ for the more stringent magnitude limit of $m_{\rm R_c}{<}22.5$. The fraction is relatively low compared to other intermediate-redshift clusters \citep[see][]{ma08,poggianti09,tran03}.

\subsection{Morphological analysis}\label{sec:morph}

Taking advantage of the availability of high-resolution imaging data obtained with HST/ACS for the central 1~Mpc region of MACSJ0025 (the sub-panel in Fig.~\ref{plane}), we attempt to examine the morphological transformation of galaxies during the cluster merger. Within this region, morphological types are assigned to all cluster members with m$_{\rm R_C}<23.0$, including 80 spectroscopically identified galaxies, as well as 28 galaxies classified as cluster members based on their photomotric redshifts, using three broad categories, namely elliptical (E), S0, and late-type (S). The classification is performed visually by examining galaxy images in the F814W band (in random order) using templates from similar classification efforts, e.g., \citet{postman05}. Fig.~\ref{example_stamps} shows postage stamps for a selected subset of galaxies with the visual morphological classification noted. To test the robustness of the visual classification and to obtain a quantitative measure of galaxy morphology, we also study the bumpiness--Sersic $n$ distribution for our galaxies, following the technique outlined in \citet{blakeslee06}.

\begin{figure*}
  \epsfxsize=0.45\textwidth 
  \epsffile{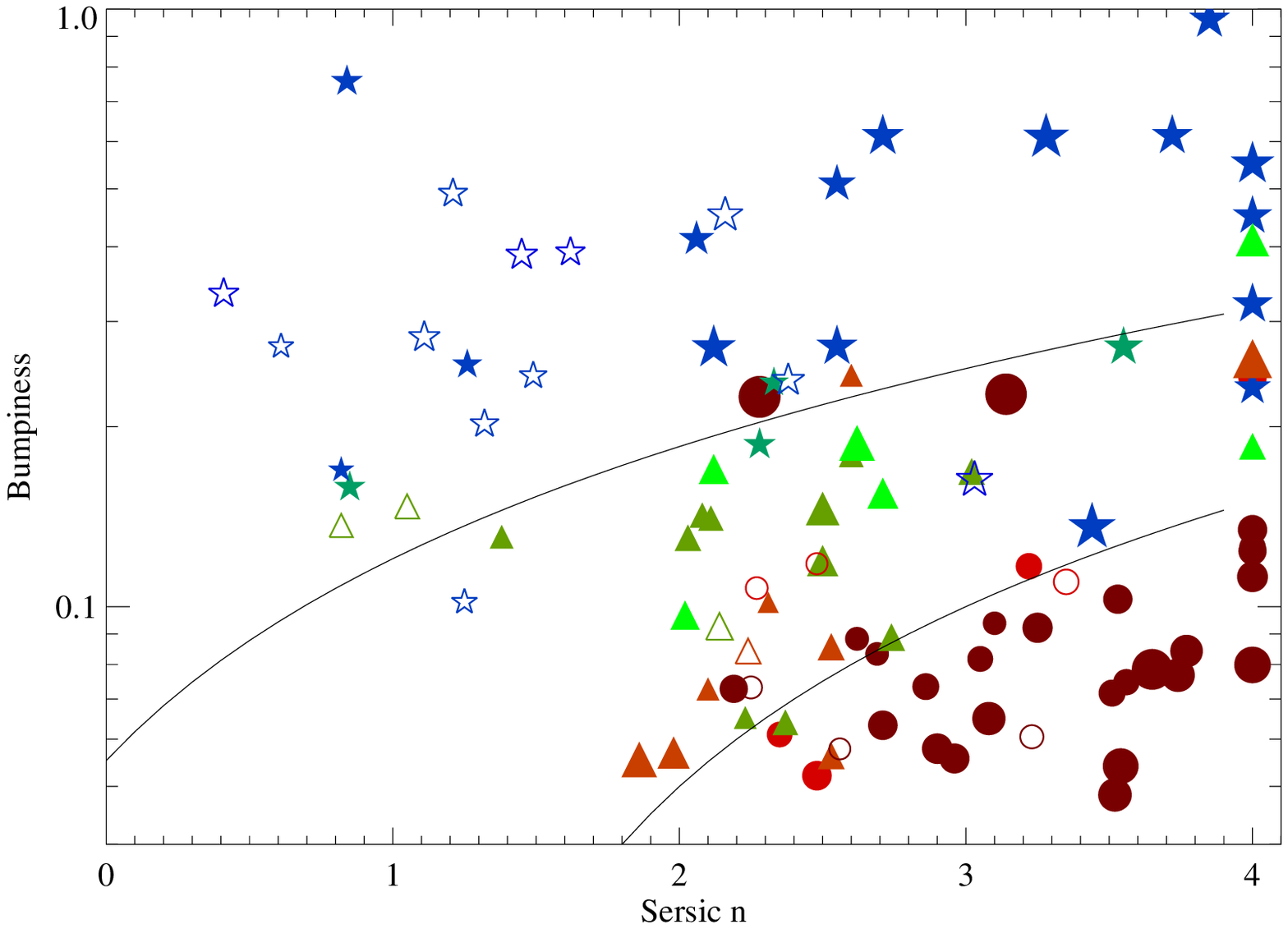}  
  \epsfxsize=0.45\textwidth 
  \epsffile{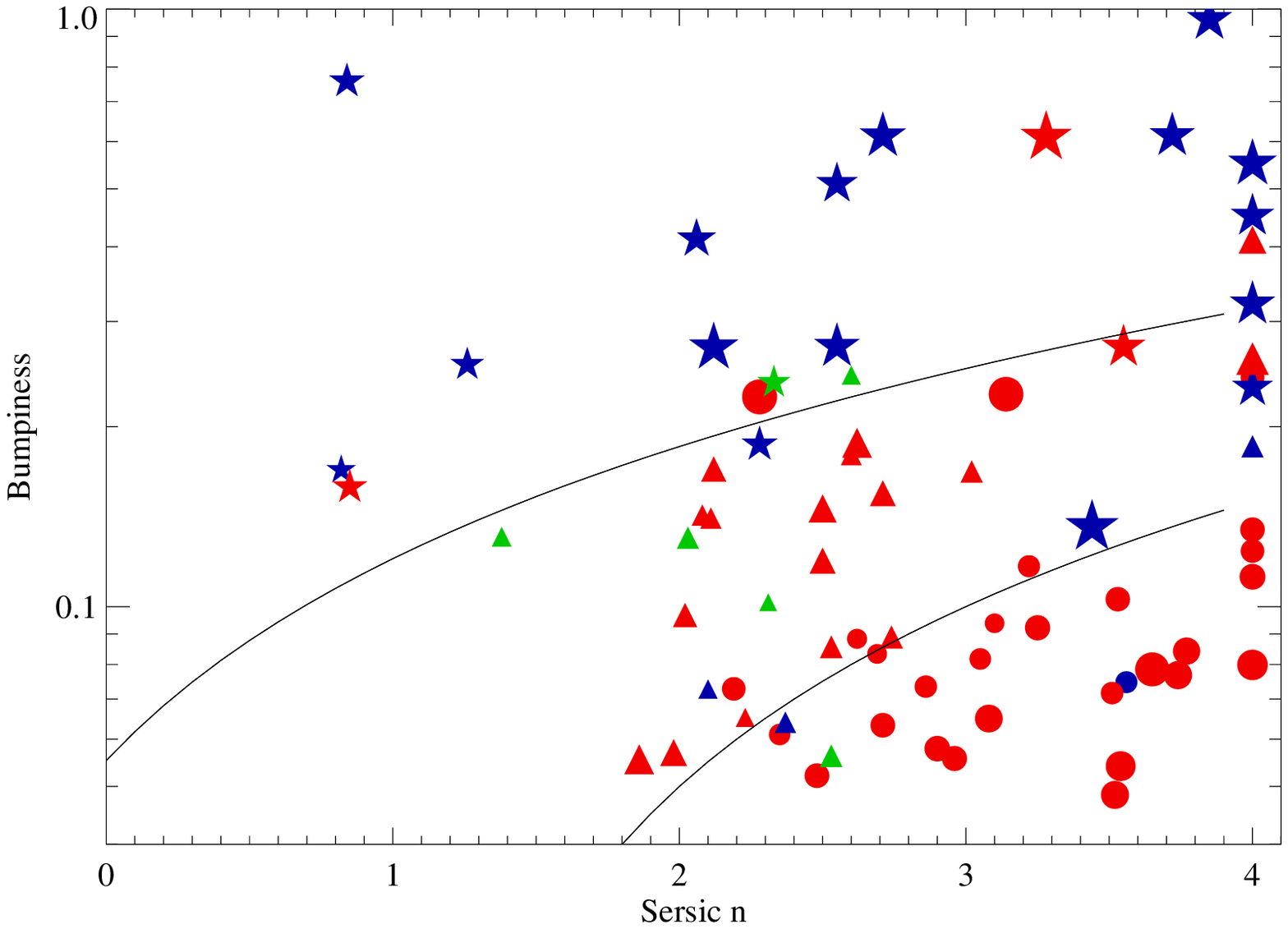}  
  \caption[f7]{Distribution of Sersic indices and bumpiness parameters of MACSJ0025 cluster members. The symbols indicate the morphological classes determined visually: early-type (circles), lenticular (S0, triangles), and late-type galaxies (stars). The symbol size is proportional to the total brightness in R$_c$. The two solid lines show the linear relations \mbox{$B=0.065(n+0.8)$} and \mbox{$B=0.05(n-1)$}, which divide early- and late-type galaxies, and S0 and elliptical galaxies, respectively \citep{blakeslee06}. In the left panel, the shading indicates morphological sub-type (black represents E, very dark grey E/S0, dark grey S0/E, gray S0, and so on). Filled symbols identify cluster members with spectroscopic redshifts; open symbols mark galaxies classified as cluster members based on their photometric redshift.  In the right panel, we show only cluster members that have spectroscopic data; here the color indicates the spectroscopic types: blue for emission-line, green for E+A, and red for absorption-line galaxies (see Section~\ref{sec:result_morph_spec}).
\label{nB}}
\end{figure*}

\subsubsection{Surface-brightness fitting}\label{sec:morph_galfit}

We fit the two-dimensional surface-brightness distribution, as obtained from ACS F814W images of our galaxies, with a Sersic model and an additional sky-background component using GALFIT\footnote{\url{http://users.obs.carnegiescience.edu/peng/work/galfit/galfit.html}}\citep{galfit02}. All parameters of the Sersic model, including the coordinates of the galaxy centre, normalization, Sersic index, effective radius, ellipticity, and orientation are fitted simultaneously. Following \citet{blakeslee06}, we constrain the Sersic index, $n$, to \mbox{$0.2<n<4.0$} since larger values do not substantially improve the fit but sometimes cause the best-fit value of the effective radius to become unreasonable large. For the initial values of the fit parameters, critical for GALFIT, we adopt numbers derived using SExtractor, except for the Sersic index, which cannot be calculated directly using SExtractor and is set to 2.5 for all galaxies. We compare and examine all fit results, including the residuals, to ensure that the algorithm did not get trapped in a local minimum. If the fit does not converge or produces implausible results, the initial parameters are adjusted manually.

GALFIT is run on images whose size is determined by 8~FLUX\_RADIUS estimated using SExtractor, but requiring a minimal size of 5$\arcsec$. A large image size is advatageous in that it provides better estimates of the sky background and the galaxy outskirts. On the other hand, larger images will also include more neighbouring objects, especially near the dense cluster core. We address this issue by masking out any close-by objects which are not connected to the target galaxy in the segmentation map generated by SExtractor. If a second galaxy is connected to the target galaxy in the section map, it is fitted with a separate Sersic model. 

Galaxies visually classified as irregular are excluded from Sersic model fits.

\subsubsection{Bumpiness}\label{sec:morph_bumpy}

The Sersic index is a well-defined parameter quantifying the concentration of the bulge; it correlates with morphology such that early-type galaxies usually have larger Sersic indices \citep{andredakis95}. However, the range of Sersic indices observed for a given morphological type is too large for it to be reliable morphological indicator in its own right. One of the reason for which Sersic indices cannot be mapped accurately to morphologial types is that, in reality, galaxies are much more complex than the smooth Sersic model: more than one Sersic model may be needed to describe the bulge and the disk, and non-radial features, such as spiral arms or tidal tails (crucial for visual classifications) cannot be modelled at all. To overcome this limitation, various alternative and additional techniques \citep{abraham96b, takamiya99, conselice03, blakeslee06} have been introduced. From these approaches, we have chosen the $B{-}n$ classification defined in \citet{blakeslee06}, because of its promising performance in distinguishing S0 galaxies from elliptical and spiral galaxies. 

Following \citet{blakeslee06}, we define the bumpiness ($B$) as the normalized residual of the surface-brightness distribution, $I$, and the fitted Sersic model, $S{(R_e,n)}$: 
\begin{equation}
B = { \sqrt{ \langle\left[\, I - S{(R_e,n)} \right]_s^2\rangle - \langle\sigma_s^2\rangle}
     \over
     \langle S{(R_e,n)}\rangle
    \label{eq:Bdef}} \,,
\end{equation}
where $R_e$ and $n$ are the effective radius and index of the Sersic model, and $\sigma$ is the flux uncertainty of the observed surface-brightness image. We compute these averages over an annulus with an inner radius of two pixels from the galaxy centre and an outer radius of $2R_e$. The subscript $s$ indicates smoothing by a Gaussian function with FWHM$=0.085\arcsec$.

 To compare the ``visual" morphology and the $B{-}n$ classification, we plot bumpiness against Sersic index for our morphological sample (left panel of Fig.~\ref{nB}). We find that the division lines of \citet{blakeslee06} are also applicable to our sample, supporting the notion of \citet{vanderwel07} that the $B{-}$n technique is robust and does not require adjustments for different studies. Several outliers are notable in Fig~\ref{nB}. Individual inspection reveals that, for instance, two galaxies visually classified as early type (red circles near the upper division line) are both Brightest Cluster Galaxies (BCGs), which cannot be fitted well by a power-law profile \citep[see e.g.][]{laine03,lauer07} resulting in $B{-}n$ parameters close to that of late-type galaxy. On the other hand, the isolated galaxy close to the lower division line (bumpiness $B\sim0.12$ and Sersic index $n\sim3.4$), visually identified as a face-on late-type spiral with faint but obvious spiral arms, exhibits a very faint disk, but a very prominent bulge, a combination which results in a small $B$ value.
   
Indeed, \citet{blakeslee06} note that the $B{-}n$ classification scheme does not always identify S0 galaxies reliably, and that the agreement with visually identified S0 is about $60\%$. This is consistent with our own findings: about $60\%$ of S0s (as identified by their location in the $B{-}n$ diagram) are also visual S0s, and about $80\%$ of visual S0a fall between the dividing lines of Fig.~\ref{nB}. We will, in the following, use both classifications to check the impact of this uncertainty on our conclusions.

\section{Results}\label{sec:result}

\begin{figure*}
  \epsfxsize=0.9\textwidth 
  \epsffile{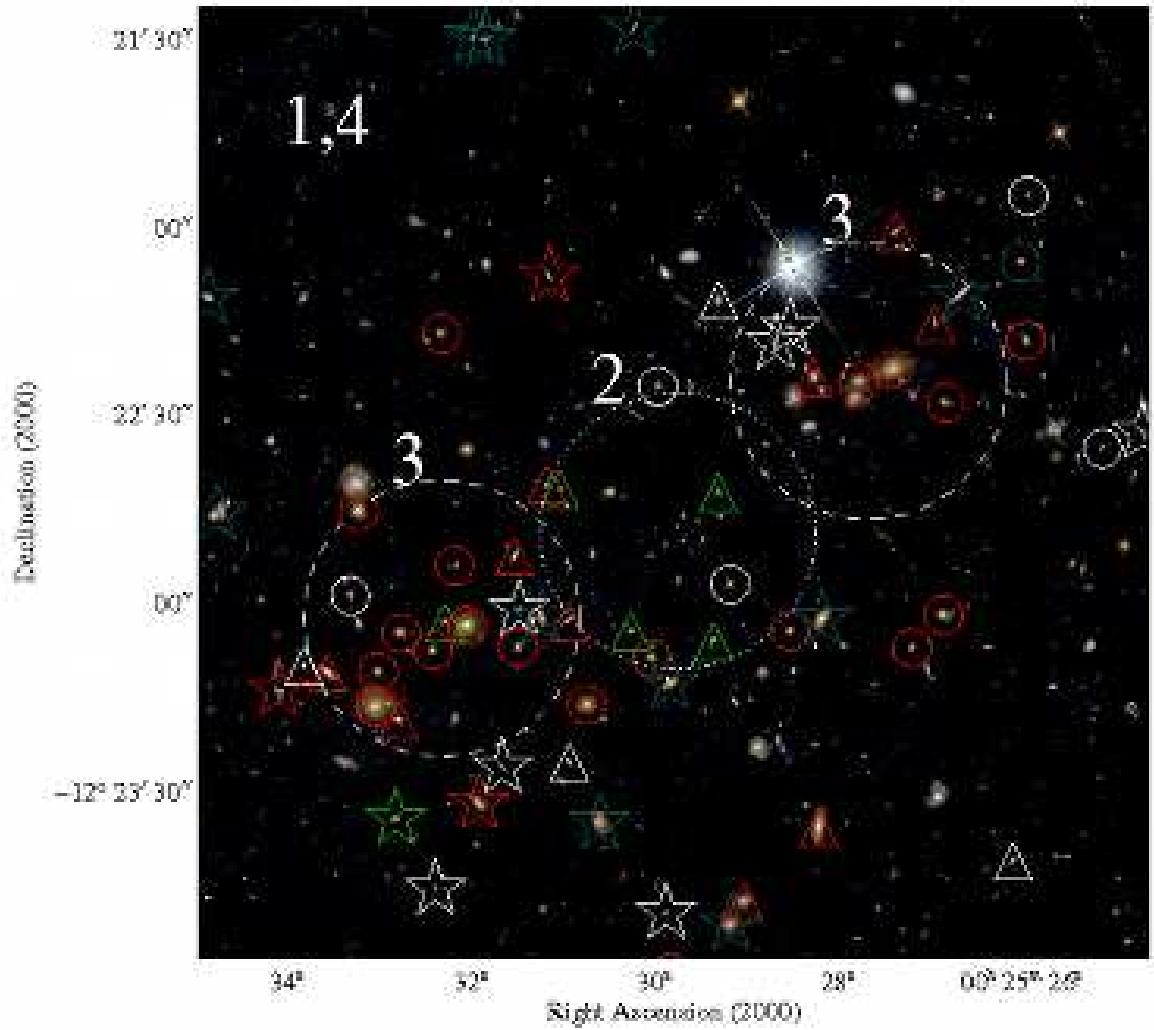} \caption[f8]{The
    distribution of galaxies of different spectral types in the cluster
    centre. Symbols are as follows: triangles mark lenticular galaxies, stars
    mark spirals, and circles mark elliptical galaxies. All symbols are also
    colour-coded, such that blue denotes emission lines in the galaxy spectrum,
    red absorption lines only, green marks E+A galaxies, and white flags
    galaxies identified as cluster members only through their photometric
    redshifts. The three large circles, finally, delineate the three
    study regions labelled as discussed in \S\ref{sec:result_morph_fraction}\label{hstcentre}.}
\end{figure*}

\subsection{Morphology} \label{sec:result_morph}

\subsubsection{Correlation between morphology and spectral type} \label{sec:result_morph_spec}

We compare the morphology of cluster members of different spectral types in the right-hand panel of Fig.~\ref{nB}. Not surprisingly, almost all of the ``visual'' late-type galaxies feature emission lines whereas almost all of the ``visual'' S0 and early-type galaxies show pure absorption spectra. Notable exceptions are three so-called passive spirals \citep[first found in][]{couch98,dressler99,poggianti99} and, conversely, one elliptical and three lenticular galaxies (again, visually classified) which exhibit emission lines in their spectra. 

Of particular interest to us, in the context of this work, are the E+A galaxies. We find them to be very similar to the population of lenticular galaxies in all respects, i.e., visual morphology (see Fig.~\ref{example_stamps}), value of the bumpiness parameter, and in the relatively narrow range in Sersic index values ($n \sim 2$) that they exhibit. Almost since their discovery in clusters \citep{couch94,dressler94}, the morphology of E+A galaxies has been a topic of particular interest. Many studies (see \cite{goto05} as well as the recent review in \cite{vergani10}) find E+A galaxies to be mostly bulge dominated, although evidence of interaction can usually be found, especially for post-starburst galaxies in the field. The findings presented by us here are consistent with earlier work \citep[e.g.\ for Coma,][]{poggianti04} in that we also find E+A galaxies to be bulge-dominated systems, although their surface-brightness profiles are significantly flatter than those of other absorption-line galaxies. Visual examination of the post-starburst galaxies in MACSJ0025 does, however, not reveal any evidence of recent interaction, a finding that is supported by the bumpiness values measured which are much lower than those of late-type galaxies. These trends (illustrated in Figs.~\ref{nB} and \ref{example_stamps}) add to a picture in which E+A galaxies in clusters exhibit different properties, and thus most likely follow different evolutionary paths, than their counterparts in the field \citep[see also][]{tran03,goto05,poggianti09}.

\begin{figure}
  \epsfxsize=0.47\textwidth 
  \epsffile{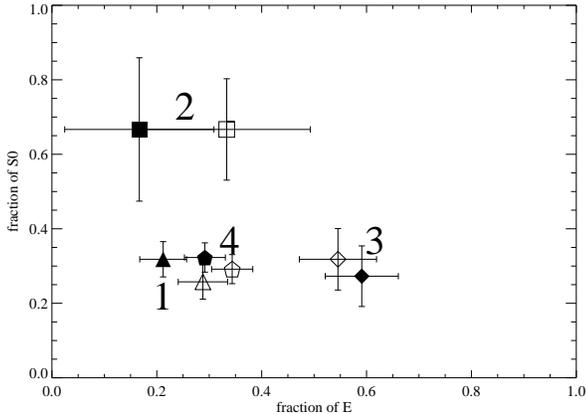}
  \caption[f9]{Fraction of early-type and lenticular clusters members in
    different regions of the clusters. The open symbols show results based on
    visual morphological classification; values marked by filled symbols were
    obtained using the B-n classification scheme. The labels identify the four
    study regions following the convention established in Fig.~\ref{hstcentre}
    and described in the text. The plotted error bars represent $1\sigma$
    confidence.  Note the high fraction of elliptical galaxies near the
    lensing-derived mass peaks (region 3), and (less statistically significant)
    the increased fraction of lenticulars near the X-ray peak in between (region
    2).\label{morph_fraction}}
\end{figure}

\subsubsection{Spatial distribution of morphological types}\label{sec:result_morph_fraction}

We calculate the fraction of morphological types in four different regions: 1) $1$~Mpc (radius) around the X-ray luminosity peak, i.e.\ essentially the entire ACS field of view, 2) $0.15$~Mpc around the X-ray luminosity peak, 3) $0.15$~Mpc around the peaks of the lensing-derived mass distribution, and 4) region (1) again but this time excluding regions (2) and (3) (see Fig.~\ref{hstcentre}). The results are summarized in Table~\ref{table_morph} and Fig.~\ref{morph_fraction}, and are consistent for our visual and the $B-n$ classification. We note the following trends: a) elliptical galaxies dominate the regions around the peaks of the mass distribution, as expected from the well known morphology-density relation \citep[][]{dressler80}; b) the region between these peaks, however, features an elevated fraction of S0s, $0.67\pm0.14$, inconsistent with the global value for region (1) at the 90\% confidence level.

\begin{figure}
  \epsfxsize=0.47\textwidth 
  \epsffile{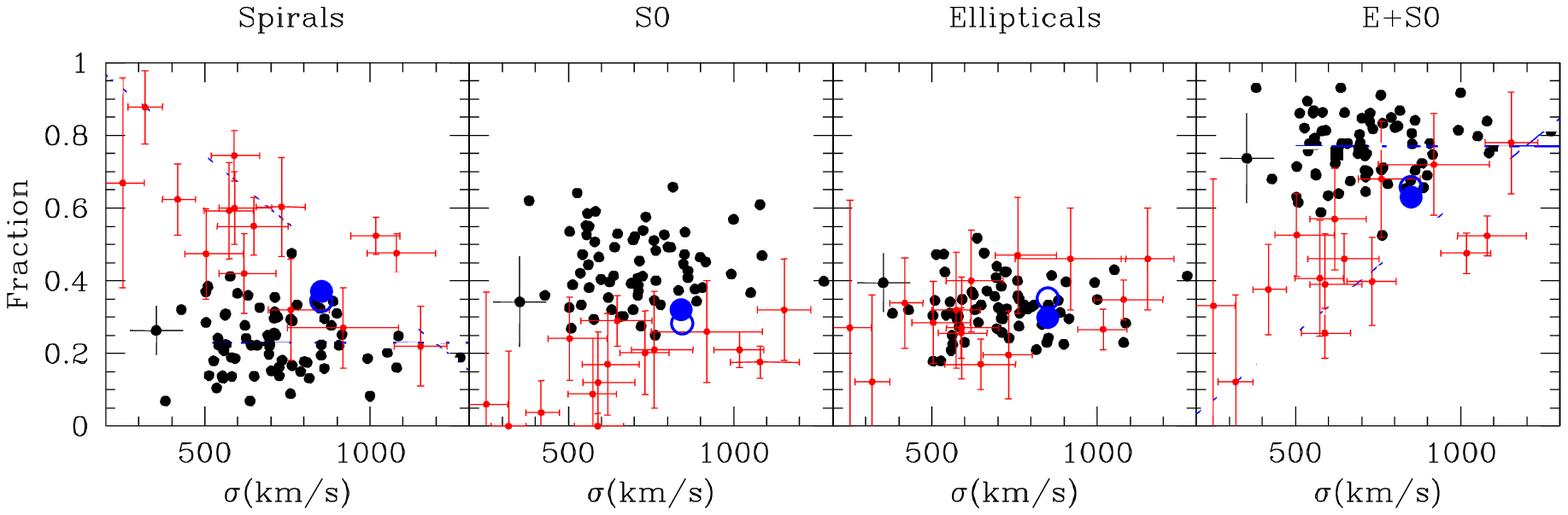}\\
  \epsfxsize=0.47\textwidth
  \epsffile{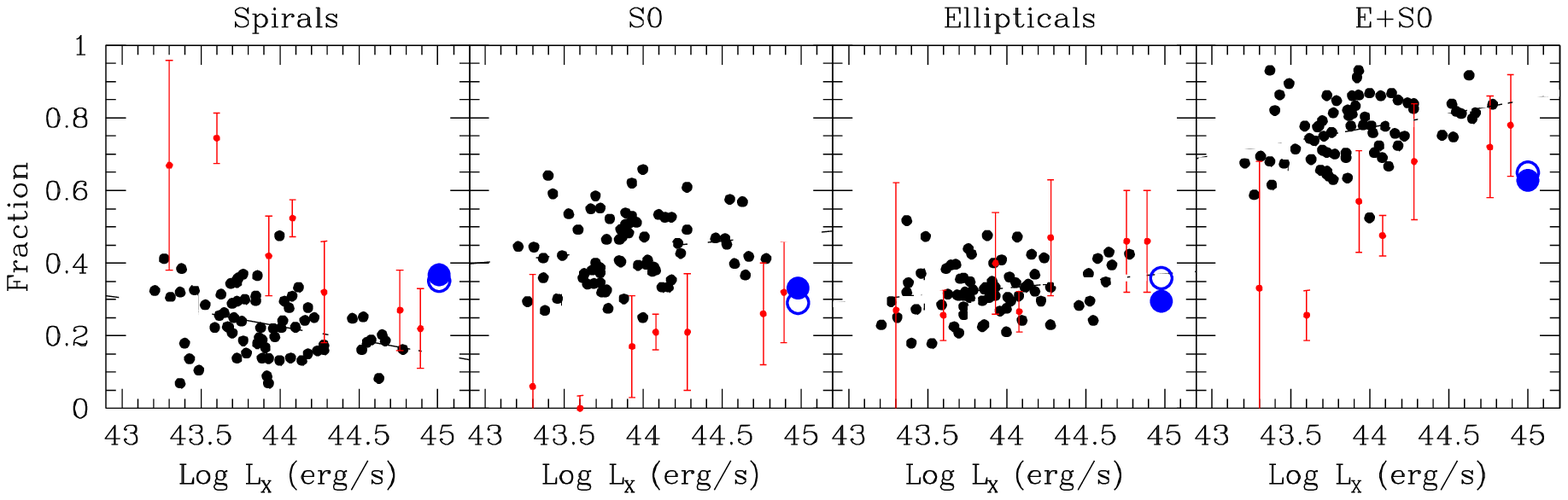}   
  \caption[f10]{Reproduced from \citet{poggianti09b}. Black circles denote nearby clusters (\mbox{$0.04<$z$<0.07$}), and red circles mark distant clusters (\mbox{$0.5<$z$<1.2$}). The open (visual classification) and filled (B-n classification) blue circles show our morphological fractions for MACSJ0025 evaluated within the same radius of $0.6$~R$_{vir}$. \label{morph_fraction_poggianti}}
\end{figure}

In general, the global fractions of the three morphological galaxy types in MACSJ0025 fall within the range of value observed in other clusters at similar redshift \citep[e.g.][]{postman05,desai07,blakeslee06,poggianti09b}. To facilitate a comparison with previous work we recompute our results for a region defined by $0.6$~R$_{vir}=0.75$~Mpc (for MACSJ0025) and a magnitude limit (M$_v < -20.0$), thereby adopting the constraints used in the study by \citet{poggianti09b}. We find our results for MACSJ0025 to, roughly, follow the trends with velocity dispersion X-ray luminosity observed for other distant clusters (see Fig.~\ref{morph_fraction_poggianti}), but note that systematic effects are likely to add to the already substantial statistical uncertainty: infall along the line of sight (a well known bias for optically selected clusters) will temporarily inflate the velocity dispersion, and merger activity along any axis will boost the X-ray luminosity beyond the virial value. With this major caveat in mind, we note nonetheless that the fraction of S and S0 in MACSJ0025 is the highest among the clusters with z$>0.5$.

%Table 3 : fraction of morphology type 
\begin{table*}
\centering
%  \tabletypesize{\scriptsize}
%  \tablewidth{0pc}
%  \tablecolumns{8} 
\caption{Fraction of morphological types\label{table_morph}}
\begin{tabular}{l|c|ccc|ccc}
\hline 
Region  & $\#$ of galaxies & \multicolumn{3}{|c}{visual}   & \multicolumn{3}{|c}{$B{-}n$ identified}  \\  
              &                           & S            &   E     &         S0  &   S  &  E  &  S0  \\
\hline
1: entire ACS image (1Mpc)                        & $96$  & $0.36\pm0.04$   & $0.34\pm0.04$   &  $0.29\pm0.04$    &  $0.39\pm0.04$   & $0.29\pm0.04$   &  $0.32\pm0.04$     \\
2: X-ray centre (0.15Mpc)					 & $6$    & $0$ & $0.33\pm0.16$ &  $0.67\pm0.14$  &  $0.17\pm0.14$ & $0.17\pm0.14$ &  $0.67\pm0.19$    \\
3: Lensing peaks (0.15Mpc)					 & $22$  & $0.14\pm0.07  $ & $0.55\pm0.07$   &  $0.32\pm0.08$    &  $0.14\pm0.07  $ & $0.59\pm0.07$   &  $0.27\pm0.08$   \\
4: Reg 1 but excluding Reg 2,3                   & $66$  & $0.44\pm0.05 $  & $0.29\pm0.05$   &  $0.27\pm0.05$    &  $0.47\pm0.04 $  & $0.21\pm0.04$   &  $0.32\pm0.05$    \\
5: X-ray centre (0.75Mpc)$^{*}$& $89$ & $0.35\pm0.04$ & $0.36\pm0.04$ & $0.29\pm0.04$ & $0.37\pm0.04$ & $0.30\pm0.04$ & $0.33\pm0.04$  \\
\hline
\end{tabular}
\medskip \\
$^*$ Following the definition in \citet{poggianti09b}: for galaxies within $0.6$~R$_{\rm vir} = 0.75$~Mpc and magnitude M$_{\rm v}<-20.0$.
\end{table*}

\begin{figure}
\epsfxsize=0.47\textwidth 
\epsffile{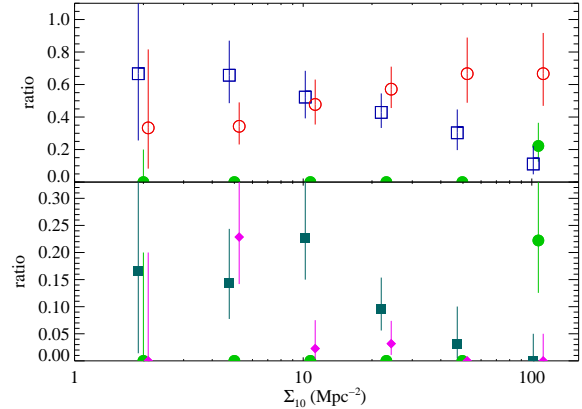}
\caption[f11]{Fraction of galaxies of different spectral
  types as a function of projected local density. The symbols denote the distribution of different spectral types: the red-circles for absorption-line galaxies, the green-circles for E+A galaxies, blue-squares for emission-line galaxies, cyan-squares for e(a), and magenta-diamonds for e(b). Error bars assume Poisson statistics. Some of the symbols are slightly shifted horizontally for clarity.\label{galxdensity}}
\end{figure}

\begin{figure}
\epsfxsize=0.47\textwidth 
\epsffile{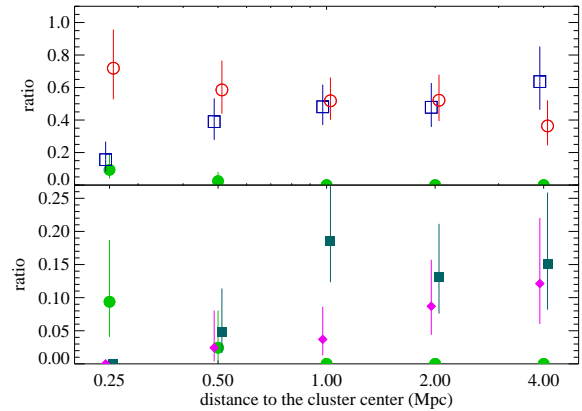}
\caption[f12]{Distribution of galaxies of different
  spectral type with cluster-centric distance. The symbols the same as in
  Fig.~\ref{galxdensity}. In each bin the fraction is calculated by dividing the
  number of member galaxies of a given spectral type by the total number of
  member galaxies observed spectroscopically. The plotted error bars assume
  Poisson statistics. Some symbols are shifted slightly by $\pm0.02$ Mpc for
  clarity. \label{galxradial}}
\end{figure}

\begin{figure*} 
\epsfxsize=0.45\textwidth 
\epsffile{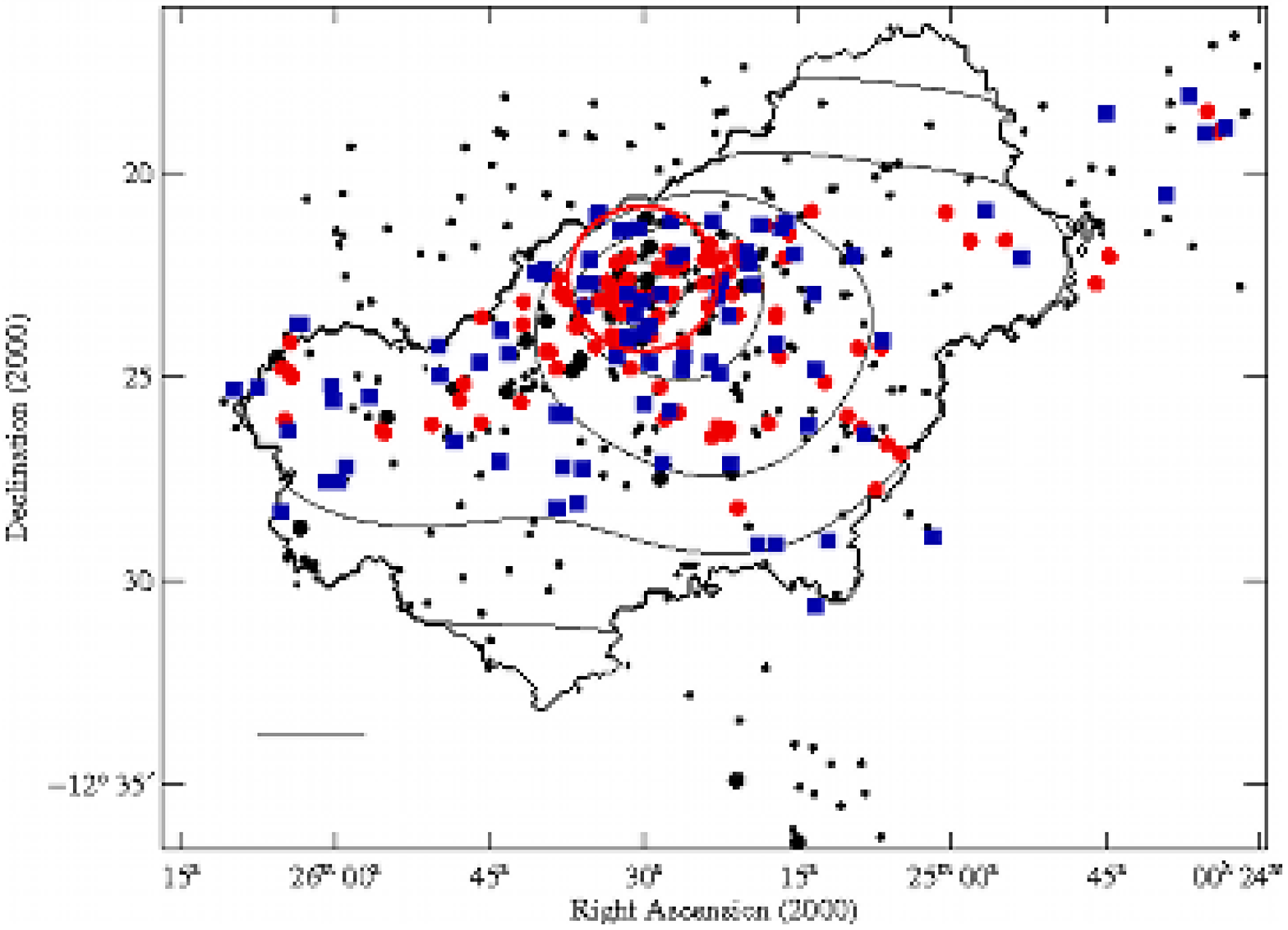}
\epsfxsize=0.45\textwidth 
\epsffile{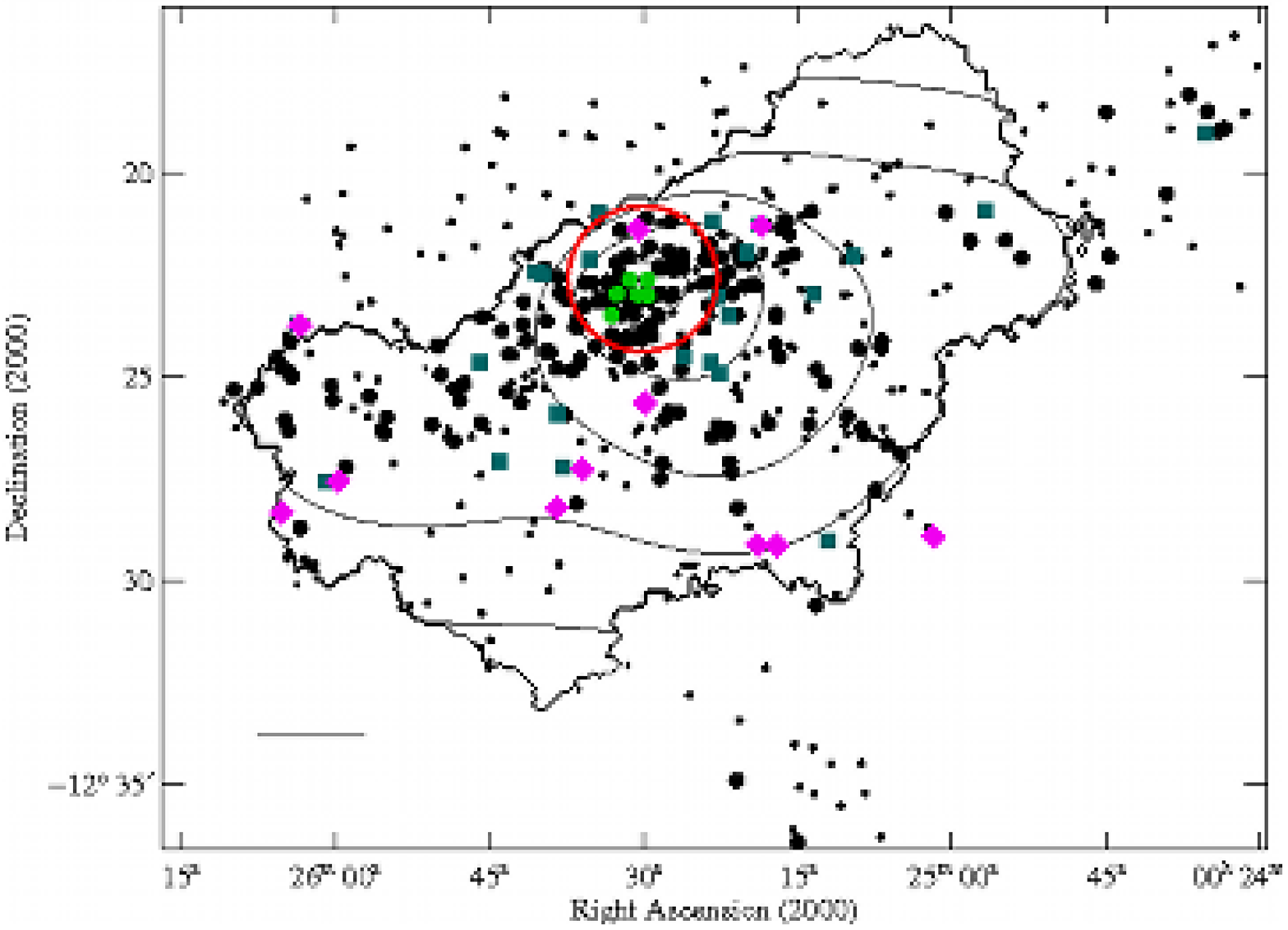}
\caption[f13]{Projected spatial distribution of
  different galaxy types in the MACSJ0025 system. The symbols are the same in Fig.~\ref{galxdensity}. In addition, black circles mark the locations of the cluster members with spectroscopic redshift and small black dots mark the locations
  of all galaxies within our study region that have been observed
  spectroscopically. Also shown are the galaxy-density contours from
  Fig.~\ref{plane}. The red circles have radii of $R_{rs}$, as estimated in \S\ref{sec:dynamics}, and a scale of 1~Mpc is drawn on the bottom-left corner of the figure.
\label{galxdistr}}
\end{figure*}

\subsection{Spatial distribution of spectral galaxy types}
\subsubsection{Projected galaxy density}\label{sec:spect_densdistr}

In Fig.~\ref{galxdensity} we plot the fraction of galaxies of different spectral
type as a function of the projected local galaxy density, $\Sigma_{10}$,
estimated using all photometrically selected cluster members.  The observed
monotonic fall and rise of the fractions of emission- and absorption-line
galaxies (top panel) with increasing galaxy density, $\Sigma_{10}$, is expected
from the relation between star-formation rate and local density, as well as from
the morphology-density relation, established by studies of other clusters at
intermediate redshift, e.g.\ \citet{dressler80, poggianti99, kodama01}. The
fraction of absorption-line galaxies in the densest regions is about $70\%$,
dropping to about $30\%$ at the lowest density studied in this
paper. Conversely, the fraction of emission-line galaxies increases to
$\sim70\%$ at the lowest density probed here, which is consistent with the fraction of
emission-line galaxies in the field \citep{gomez03,cooper07}.

In the bottom panel of Fig.~\ref{galxdensity}, we see that all six E+A galaxies identified in MACSJ0025 reside in regions of very high galaxy density. The distribution of E+A galaxies in MACSJ0025 is thus very different from that observed in other intermediate-redshift clusters \citep[e.g. ][]{dressler99,tran03,ma09} where E+A galaxies are found in a wide range of projected galaxy densities and do not favor the densest environment. 

In contrast to the distribution of E+A galaxies, the fraction of e(a) galaxies is the lowest in the densest regions of the cluster, increases to a maximum at ${\Sigma}_{10}\sim10$~Mpc$^{-2}$ (which happens to be the density at which the fractions of emission- and absorption-line galaxies are the same), but does not rise further at lower densities. This result is consistent with the consensus that intermediate galaxy densities provide the most efficient environment for star formation.

\subsubsection{Cluster-centric distance}\label{sec:spect_spatdistr}

Plotting the distribution of cluster galaxies of different spectral type as a function of cluster-centric distance (Fig.~\ref{galxradial}, a two-dimensional version is shown in Fig.~\ref{galxdistr}), we find a trend that is very similar to the one shown in Fig.~\ref{galxdensity}, i.e.\ a smooth monotonic decrease/increase of the absorption/emission-line galaxies with increasing distance from the cluster centre. Note, however, that the fraction of emission-line galaxies is significantly suppressed only within $r\sim1$\,Mpc, which is approximately the size of the virial radius\footnote{Since the cluster-centric radius is difficult to define in this merging system we adopted, for this analysis, the location of the peak of the X-ray surface brightness as the cluster centre. The exact choice is not critical though at the coarse binning used in Fig.~\ref{galxradial}. We confirmed that our results do not change when the cluster-centric radius is computed using a more complicated elliptical model where the two focal points are given by the two peaks of the lensing mass distribution.}  

Similar to the distribution as a function of galaxy density depicted in Fig.~\ref{galxdensity}, the distribution of e(a) galaxies as a function of cluster-centric distance also peaks at the scale ($r\sim1$\,Mpc) at which the fractions of emission- and absorption-line galaxies are roughly equal, with no e(a) galaxies being found in the very centre of the cluster. The trend can be seen more clearly in the two-dimensional distribution shown in the right panel of Fig.~\ref{galxdistr}. This result implies the presence of an efficient trigger of starburst activity near the virial radius of the cluster. 

Most interestingly, however, all of Figs.~\ref{hstcentre}, \ref{galxradial}, and \ref{galxdistr} illustrate clearly that E+A galaxies appear to reside preferentially in the very centre region of the cluster (note: we define the cluster centre to be the location of the peak of the X-ray emission), a result that is in obvious conflict with the consensus that E+A galaxies in clusters avoid the very cluster core \citep[see][and references therein]{ma08}. As we shall discuss in Section~\ref{sec:discussion} this apparent conflict can be resolved once the geometry and dynamics of this merger are taken into account.

\subsubsection{Redshift distribution}\label{sec:spect_zdistr}

\begin{figure}
\epsfxsize=0.47\textwidth
\epsffile{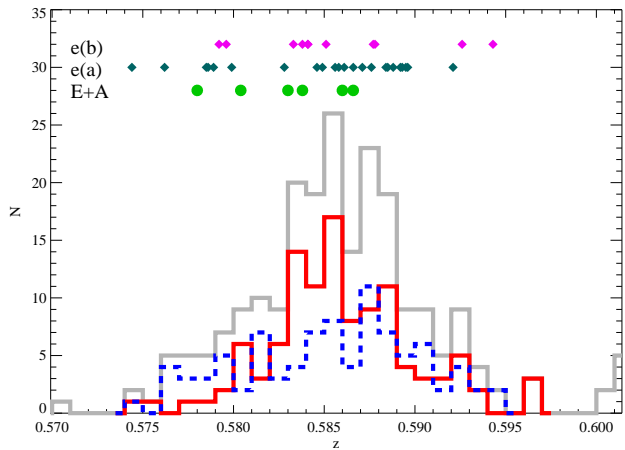}
\caption[f14]{Redshift distribution of different spectral types of cluster members. Redshift distribution of the emission-line and absorption-line cluster members are shown, respectively, with blue and red histograms, overlaying on the redshift distribution of the entire spectrum sample (grey).  The magenta, cyan, green points, mark the redshift of the e(b), e(a) and E+A galaxies. \label{spect_zdistr}}
\end{figure}

The redshift distribution for different spectral types of cluster members is shown in Fig.~\ref{spect_zdistr}. It is narrower for absorption-line galaxies than for emission-line galaxies, which is expected if the infalling population is dominated by emission-line galaxies. More importantly, we note that the redshift distributions of E+A, e(a) and e(b) galaxies are not significantly different from that of emission- or absorption-line galaxies, consistent with the hypothesis that, in a dynamical sense, they are drawn from the same parent population.

\subsection{Color--magnitude distribution of different spectral types}\label{sec:spect_cmd}

\begin{figure}
  \epsfxsize=0.47\textwidth 
  \epsffile{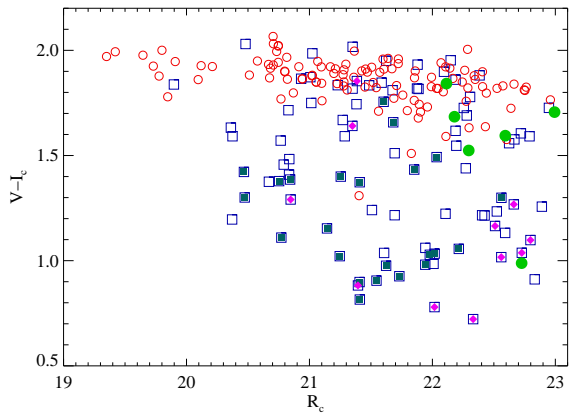}  
  \caption[f15]{Color-magnitude diagram for galaxies of
    different spectral type. The the symbols are the same as in Fig.~\ref{galxdensity}. The E+A galaxies (green filled circles) are fainter than $\rm m_{R_c}>22.0$ and are located mostly in the ``green valley" below the red-sequence marked by the absorption-line galaxies (red open circles). Compare to the E+A galaxies, the e(a) galaxies (cyan-squares) are brighter and exhibit a broad range of color in ``blue cloud". \label{cc}}
\end{figure}

We plot the V--I$_c$ colour and R$_c$ magnitude of all cluster members in Fig.~\ref{cc}, using different symbols to denote spectral type. Similar to previous results \citep[e.g.][]{ma08, poggianti09b, tran03}, the E+A galaxies are found to inhabit the ``green valley", with one exception, for which the (groundbased) colour measurement is erroneous due to contamination by a blue gravitational arc, as revealed in our ACS image (top left image in Fig.~\ref{example_stamps}). All E+A cluster members are fainter than M$_R >-20.7$, unlike the E+A galaxies in other intermediate redshift clusters, including MACSJ0717.5+3745 for which \citet{ma08} find M$_R=[-23.5,-20.0]$. The magnitudes of E+A galaxies in MACSJ0025 are thus closer to those of the faint populations of E+A galaxies found to dominate in local clusters, e.g.\ Coma \citep{poggianti04} and Abell~3921 \citep{ferrari05}. In the B band, the range in absolute magnitude of these E+A galaxies is M$_B=[-19.3,-20.2]$, which is just above the limit of the typical magnitude of early-type dwarf galaxies, $-19<{\rm M}_B<-16$ \citep[e.g. ][]{barazza09}, found in local galaxy clusters and groups \citep[see ][and references therein]{binggeli90, mateo98, haines07,barazza09}. Note, however, that the B-band magnitudes of the E+A galaxies in MACSJ0025 are boosted by the presence of a young stellar population, and that their brightness will drop by 3--4 magnitudes\footnote{Estimated using the simple stellar synthesis model from \citet{BC03}}, as they evolve passively from $z=0.59$. Moreover, the morphology of the E+A galaxies in MACSJ0025 is well matched to that of early-type dwarf galaxies which are characterized by a flatter surface-brightness profile (Sersic index of about 2; \citep[e.g. ][]{smith09}) than for typical ellipticals.

Unlike E+A galaxies, e(a) galaxies exhibit a broad range of magnitudes within the ``blue cloud". Because of their strong H$_{\delta}$ absorption features and the moderate EW of their emission lines, e(a) galaxies are believed to be possible progenitors of E+A galaxies\citep{poggianti99}. They are, in general, brighter than E+A galaxies, which might, again, be due to excess stellar light during the starburst phase. In this scenario, their magnitude would drop to the levels typical of E+A galaxies once the starburst is terminated.

Finally, we note that e(b) galaxies are observed predominantly at the very blue end of the explored colour range. We therefore suspect that many cluster members of e(b) type are missed in our study due to the colour criterion applied for the selection of the spectroscopic sample; hence, the fraction of e(b) galaxies should be considered as a lower limit.

\section{Physical interpretation}\label{sec:discussion}

\subsection{A simple model}\label{sec:toymodel}

Because of its simple merger geometry (major equal-mass merger, in the plane of the sky, after first core passage), MACSJ0025 provides us with an excellent opportunities to study the effect of cluster mergers on fundamental galaxy properties, specifically their morphology and star formation activity.

\begin{figure}
  \epsfxsize=0.47\textwidth 
  \epsffile{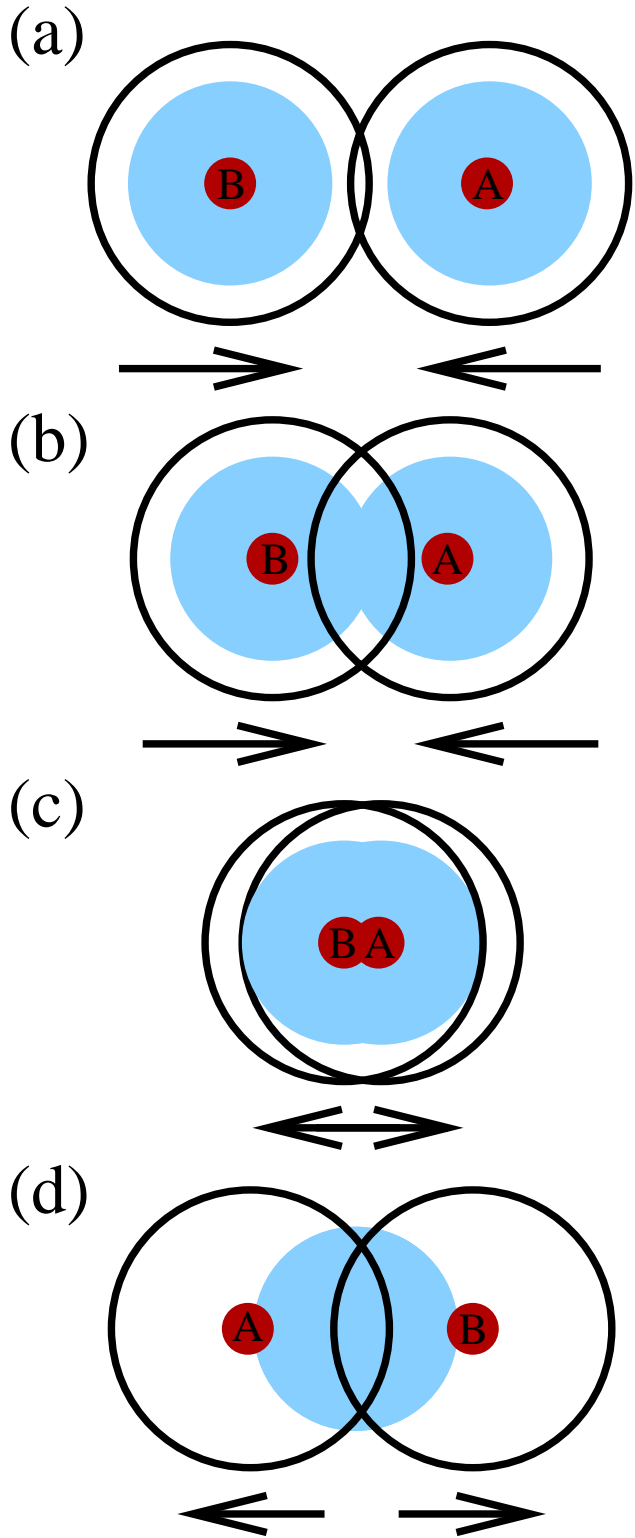}  
  \caption[f16]{A schematic illustration of the collision process. Red circles mark the centres of the dark-matter distributions of the two clusters, while light blue and black circles represent the distribution of the ICM and the cluster galaxy population, respectively. See text for discussion. \label{toymodel}}
\end{figure}

Our understanding of the merger history and dynamics of MACSJ0025 is depicted schematically in Fig.~\ref{toymodel}. The key phases and physical processes at work can be described as follows:

\begin{description}
\item[Phase A:] pre-contact; no significant galaxy-gas interaction yet; relative galaxy velocities too high for galaxy mergers, but increasing probability of harassment events 
\item[Phase B:] first contact; increased ICM interaction induces shocks; onset of segregation between dark and luminous matter; ram-pressure stripping and tidal interactions begin as galaxies encounter steep gradients in gravitational potential and ICM density; morphological tranformation and triggered star formation
\item[Phase C:] core passage; significant ICM interaction and shock heating; non-collisional components (dark matter, galaxies) proceed as gas is left behind; galaxies on the leading edge of either cluster now fully stripped of their gas content; galaxies at the trailing edge experience Phase B effects
\item[Phase D:] present stage; dark matter cores fully separated but single ICM component concentrated at global centre of mass; gas-galaxy interactions (triggered star formation, stripping) continue as trailing galaxy populations pass through core.
\end{description}

Qualitatively, this picture is supported by the result of numerical simulation of ram-pressure stripping \citep[e.g.][]{fujita99b,kronberger08,bekki09} and  tidal interaction \citep{byrd90, bekki99, gnedin03a, gnedin03b}. Although the environments encountered in this process by galaxies at the leading and trailing edges of the individual clusters are roughly symmetrical, galaxies on the far side of the merger experience a hostile environment for longer since they encounter strong tidal fields (passage through the dark matter core) and ram-pressure stripping sequentially; in addition, they pass through the more massive and denser post-collision gaseous core. 

We note that the velocities of galaxies relative to their respective cluster core are not negligible, and hence their observed relative locations may differ substantially from the ones prior to collision. Exceptions are, however, the E+A and the e(a) populations, both of which will remain reasonably faithful tracers of environment thanks to the relatively short duration of the star-formation and quenching processes.

In the following, we will examine whether the features observed in the galaxy population of MACSJ0025 support the picture described above.

\subsection{Distribution of E+A galaxies}

As discussed in \S\ref{sec:spect_densdistr} and \S\ref{sec:spect_spatdistr}, we find all but one of the E+A galaxies in MACSJ0025 to be located near the peak of the X-ray surface brightness, and thus between the dark-matter cores of the individual clusters. This is exactly what would be expected in the scenario outlined in \S\ref{sec:toymodel} and depicted in Fig.~\ref{toymodel}. The merging process in MACSJ0025 thus appears to cause dramatic and rapid galaxy evolution (the triggering and termination of startbusts) in an environment where it is rarely observed, namely the region of highest gas density in the very cluster core.

We should mention in this context that E+A galaxies have in fact been detected in cluster cores before. In all cases, however, projection effects seem to play a major role. Specifically, the group of starburst galaxies found in Abell~851 \citep[see][and references there in]{oemler09} forms a group of high peculiar velocity falling into the cluster along the line of sight. By contrast, no evidence of projection effects is found for MACSJ0025 (see \S\ref{sec:spect_zdistr}).

\subsection{Morphological transformation}

In the merger scenario discussed above, the galaxy populations of both clusters should have been exposed to conditions conductive to trigger star formation and ram-pressure stripping for a large fraction of the duration of the cluster collision. As a result, one would expect the morphological transformation from late-type to S0 galaxies to have been particularly efficient in this system in recent times. Indeed, we measure a high global S0 fraction of about $0.3$ (Fig.~\ref{morph_fraction} and Table~\ref{table_morph}), among the highest found to date in any cluster at $z>0.5$, but still within the observed range of 0.05--0.33 \citep{poggianti09b}. Furthermore, the high global S fraction of about $0.39$ may reflect that cluster merger help the mix of late-type galaxies at outskirt. A more dramatic increase in the global S0 and S fraction would have been surprising, given that the morphological mix of galaxies in clusters is constantly driven to a global average as S0s continue to evolve into ellipticals, and late-type galaxies are being accreted from the field. A snapshot of the merger-driven transformation process is, however, provided by our view of the core of MACSJ0025 where we find all E+A galaxies to be lenticulars, leading to a local S0 fraction of about 70\%.

Our hypothesis that interactions with the gravitational potential and the intracluster medium are primarily responsible for the accelerated evolution of galaxies in this massive merger is also supported by a comparison with the morphology of E+A galaxies in the field. Signature of tidal destruction or galaxy merging can be removed efficiently and quickly by both ram pressure  \citep{kapferer08}  and tidal forces \citep{bekki01c} in the cluster core.  
%Unlike their counterparts in the field, the E+A galaxies in MACSJ0025 (but also in Coma) show no evidence of tidal destruction and galaxy merging, which is consistent with the results from simulations that both ram pressure \citep{kapferer08} and tidal forces \citep{bekki01c} in the cluster core remove morphological detail very efficiently and quickly. 

\begin{figure}
  \epsfxsize=0.47\textwidth 
  \epsffile{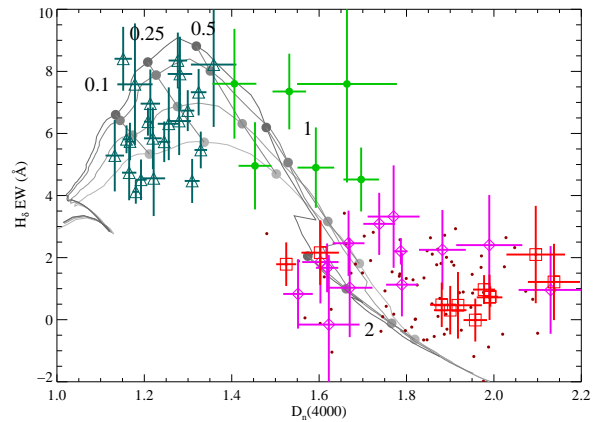}  
  \caption[f17]{The D$_n$(4000) and EW of H$\delta$ of galaxies in the cluster; cyan for e(a) galaxies, green for E+A galaxies, magenta for absorption-line galaxies in the regions farther from the cluster cores along merger axis, and red for absorption-line galaxies in region 3 of Fig.~\ref{hstcentre}. The four curves represent stellar-population synthesis models from \citet{BC03}, assuming a delta-function starburst in a 5~Gyr-year-old galaxy with exponentially decaying SFR ($\tau=1Gyr$). From top to bottom, the curves are for models with starburst fractions of $100\%$, $30\%$, $10\%$, and $5\%$. The labels along the curves denote the age of the burst in Gyr. \label{D4000}}
\end{figure}

\subsection{Temporal evolution}

We can use estimates of the duration of characteristic phases of galaxy evolution derived from spectral diagnostics for a consistency check of our simple merger model (Fig.~\ref{toymodel}).

Specifically, stellar synthesis models \citep{BC03} predict how the strength of the D$_n$(4000) break and the EW of H$_{\delta}$ vary along an evolutionary track from the beginning of the starburst to several Gyrs after. In Fig.~\ref{D4000}, the data for galaxies of the various spectral types observed in MACSJ0025 are overlaid on the theoretically expected tracks\footnote{The models assume a delta-function starburst in a 5~Gyr-year-old galaxy with exponentially decaying SFR ($\tau=1Gyr$), Solar metalicity, and a Salpeter initial mass function.}. We note that these models can only be used to describe the absorption features in the stellar spectra; emission-lines from the interstellar medium are not included. While this makes the model predictions less meaningful for emission-line galaxies, they should be applicable to the E+A galaxies that we are primarily interested in. 

 Along the track, the magnitude-weighted stellar age \citep{kauffmann03} from the models agrees qualitatively with the location of the various galaxy subpopulations. If the E+A population observed near the peak of the X-ray emission was created by the equal-mass merger in MACSJ0025, these galaxies' approximate age of 0.5--1 Gyr  (Fig.~\ref{D4000}) would provide a constraint on the time elapsed since core passage (Phase C). As interestingly, in the context of our merger model, is that the regions farther from the cluster core along the merger axis are dominated by young absorption-line galaxies in which the star formation ended about $1-2$~Gyrs ago. A high fraction of young passively evolving galaxies along the merger path is consistent with our model in which the transformation of these galaxies would have begun in phase A, i.e.\ at the onset of the cluster collision. These time scales are consistent with the predictions from numerical simulations \citep{barnes01} which show that the relative velocity of the two free-falling dark-matter halos is significantly reduced by dynamical friction during a head-on merger. 

For reference, Fig.~\ref{D4000} also shows the location of e(a) galaxies to demonstrate, qualitatively, that their stellar ages derived from this H$_{\delta}$-D$_n$(4000) method are younger than those of the E+A galaxies. A more quantitative analysis of the e(a) galaxies' locus in this graph would require more complex models which are beyond the scope of this paper.

Finally, and not surprisingly, we find the absorption-line galaxies in the two sub-cluster core (red symbols in Fig.~\ref{D4000}) to be the oldest population, consistent with the expectations that the star-formation epoch in elliptical galaxies at center of ``proto''-clusters ends  as early as z$\sim 2$\citep[e.g. ][]{blakeslee06} . 

 \subsection{Relation between starburst and cluster merger}

Simulations suggest increased star-formation activity during cluster mergers \citep[e.g.][]{bekki99}. 
However, most of the observational results \citep[e.g.][]{marcillac07} for starburst galaxies in clusters are interpreted as  being related to infall. 

The spatial distribution of e(a) galaxies in MACSJ0025 (\S\ref{sec:spect_spatdistr} and \S\ref{sec:spect_densdistr}) seems to support the infall scenario, in that most of the starburst galaxies are located at the cluster outskirts. This is consistent with the view that starbursts in e(a) galaxies are caused by galaxy-galaxy interactions which are highly improbable at the high relative velocities of galaxies closer to the cluster core. 

Nevertheless, evidence of starbursts near the X-ray centre is observed in the form of the quenched starbursts in E+A galaxies. The physical mechanism that caused these bursts is likely to be the same that ultimately quenches them, creating the E+A phase of galaxy evolution, namely the violent interaction with the dense intracluster medium encountered during the merger (b and c, in Fig.~\ref{toymodel}). The fact that no starburst or post-starburst galaxies are found in the regions farther from the cluster core, but still well within the viral radius, suggests that extreme gas densities are required to trigger bursts, which is consistent with the few instances in which such dramatic gas-galaxy encounters have actually been observed (Cortese et al.\ 2007; Richard et al.\ 2010). This still leaves us with the question, however, of why no active starbursts are presently observed near or within the newly formed gaseous core (phase d of Fig.~\ref{toymodel}). We argue that galaxies from either cluster approaching the core along the merger axis, and thus encountering a similarly hostile environment as the galaxies at the leading edge of either system during first approach, do in fact experience bursts of star formation and will, ultimately, also go through an E+A phase. Their density, however, and thus the number of such ICM-triggered starbursts decreases rapidly as the merger proceeds and only galaxies from the increasingly sparsely populated outer cluster regions continue to approach.

\section{Summary}\label{sec:summary}

\begin{itemize}

\item We present results of a extensive morphological, spectroscopic, and photometric study of the galaxy population of MACS\,J0025.4$-$1225, an equal-mass cluster merger, based on 212 galaxy spectra obtained with Keck DEIMOS as well as colour imaging of the cluster core with HST/ACS and wide-field imaging in seven passbands from Subaru SuprimeCam and CFHT/Megacam and CFHT/WIRCAM observations.

\item Six E+A galaxies are found in MACSJ0025, all of them near the cluster center defined by the X-ray surface brightness peak, in contrast to other studies which find the E+A population to avoid the cluster core. The color of these E+A galaxies fall in between those of red-sequence galaxies and emission-line galaxies. Compared to E+A galaxies in other clusters at intermediate redshift, the luminosity function of E+A galaxies in MACSJ0025 appears truncated (no galaxies with M$_R < -20.7$). Their morphology is broadly lenticular, i.e., characterized by a dominant bulge and relatively flat surface-brightness profiles (Sersic index$\sim 2$). In addition, no evidence of tidal features or other asymmetric features is found. 

\item The fraction of absorption(emission)-line galaxies is a smooth increasing (decreasing) function of projected galaxy density. A similar correlation is found with distance to the cluster center. The fraction of dusty-starburst galaxies (e(a)) is dramatically lower inside the virial radius than outside.

\item The global fractions of morphological types (E, S, S0) of galaxies in MACSJ0025 are consistent with those in clusters at $z>0.5$ compiled by \citet{poggianti09b}, although the fraction of S0 and S types observed in MACSJ0025 are the highest among the sample. While the global S0 fraction is thus not dramatically enhanced, we find the local value boosted around the X-ray surface brightness, coinciding with the excess of E+A galaxies.

\item We propose a simple model in which star-formation activity and morphological transformations are caused by dynamic interactions enhanced and accelerated by the cluster merger. In this picture, the time since the termination of the starburst phase in the observed E+A galaxies naturally matches the time elapsed since the core passage of the merger; about 0.5--1.0 Gyrs ago in the case of MACSJ0025. Additional evidence for the causal role of the merger is provided by the relatively young age of the absorption-line cluster members near the leading edges of the two clusters engaged in the collision. 

\end{itemize}

\section*{Acknowledgments}

The work of PJM was supported in part by the U.S. Department of Energy under contract number DE-AC02-76SF00515, and by the TABASGO and Kavli foundations in the form research fellowships. The HAGGLeS legacy archive project was supported by HST grant 10676. TS acknowledges support from NWO. The work is base in part on data collected at Subaru Telescope, which is operated by the National Astronomical Observatory of Japan. It is also based, in part, on observations obtained with MegaPrime/MegaCam, a joint project of CFHT and CEA/DAPNIA, at the Canada-France-Hawaii Telescope (CFHT) which is operated by the National Research Council (NRC) of Canada, the Institute National des Sciences de l'Univers of the Centre National de la Recherche Scientifique of France, and the University of Hawaii. It is also based on observations obtained with WIRCam, a joint project of CFHT, Taiwan, Korea, Canada, France, at the Canada-France-Hawaii Telescope (CFHT) which is operated by the National Research Council (NRC) of Canada, the Institute National des Sciences de l'Univers of the Centre National de la Recherche Scientifique of France, and the University of Hawaii. Again, it is based, in part, on observations with the NASA/ESA Hubble Space Telescope, obtained at the Space Telescope Science Institute, which is operated by AURA Inc, under NASA contract NAS 5-26555. The spectroscopic data presented herein were obtained at the W.M.\ Keck Observatory, which is operated as a scientific partnership among the California Institute of Technology, the University of California and the National Aeronautics and Space Administration. The Observatory was made possible by the generous financial support of the W.M. Keck Foundation.

%{\bibliographystyle{mn2e}}
%{\bibliography{m0025_paper}}

\end{document}